\documentclass[aps,twocolumn,showpacs,preprintnumbers,amsmath,floatfix,superscriptaddress,longbibliography]{revtex4-1}

\usepackage{graphicx}
\usepackage{epsfig}
\usepackage{dcolumn}
\usepackage{bbm}
\usepackage[colorlinks,linkcolor=red,anchorcolor=green,citecolor=blue,CJKbookmarks=True]{hyperref}
\usepackage{braket}
\begin{document}
\title{\texorpdfstring{$1^{-+}$}{1-+} hybrid in \texorpdfstring{$J/\psi$}{J/psi} radiative decays from lattice QCD}
\author{\small Feiyu Chen}
\email{chenfy@ihep.ac.cn}
\affiliation{\small Institute of High Energy Physics, Chinese Academy of Sciences, Beijing 100049, People's Republic of China}
\affiliation{\small School of Physics, University of Chinese Academy of Sciences, Beijing 100049, People's Republic of China}

\author{\small Xiangyu Jiang}
\email{jiangxiangyu@ihep.ac.cn}
\affiliation{\small Institute of High Energy Physics, Chinese Academy of Sciences, Beijing 100049, People's Republic of China}
\affiliation{\small School of Physics, University of Chinese Academy of Sciences, Beijing 100049, People's Republic of China}

\author{\small Ying Chen}
\email{cheny@ihep.ac.cn}
\affiliation{\small Institute of High Energy Physics, Chinese Academy of Sciences, Beijing 100049, People's Republic of China}
\affiliation{\small School of Physics, University of Chinese Academy of Sciences, Beijing 100049, People's Republic of  China}

\author{\small Ming Gong}
\affiliation{\small Institute of High Energy Physics, Chinese Academy of Sciences, Beijing 100049, People's Republic of  China}
\affiliation{\small School of Physics, University of Chinese Academy of Sciences, Beijing 100049, People's Republic of  China}

\author{\small Zhaofeng Liu}
\affiliation{\small Institute of High Energy Physics, Chinese Academy of Sciences, Beijing 100049, People's Republic of  China}
\affiliation{\small School of Physics, University of Chinese Academy of Sciences, Beijing 100049, People's Republic of  China}
\affiliation{\small Center for High Energy Physics, Peking University, Beijing 100871, People's Republic of  China}

\author{\small Chunjiang Shi}
\affiliation{\small Institute of High Energy Physics, Chinese Academy of Sciences, Beijing 100049, People's Republic of  China}
\affiliation{\small School of Physics, University of Chinese Academy of Sciences, Beijing 100049, People's Republic of  China}

\author{Wei Sun}
\affiliation{\small Institute of High Energy Physics, Chinese Academy of Sciences, Beijing 100049, People's Republic of  China}

\def\modified#1{\red{#1}}
\begin{abstract}
We present the first theoretical prediction of the partial decay width of the process $J/\psi\to \gamma \eta_1$, where $\eta_1$ is the lightest flavor singlet $1^{-+}$ hybrid meson. Our $N_f=2$ lattice QCD calculation at $m_\pi\approx 350$ MeV results in the $\eta_1$ mass $m_{\eta_1}=2.23(4)$ GeV and the related electromagnetic form factors $M_1(0)=-4.73(74)$ MeV, $E_2(0)=1.18(22)$ MeV, which give $\Gamma(J/\psi\to \gamma\eta_1)=2.04(61)$ eV. These form factors can be applied to the physical $N_f=3$ case where there should be two hybird mass eigenstates $\eta_1^{(l)}$ and $\eta_1^{(h)}$ due to the singlet-octet mixing. It is shown that the ratio of the branching fractions $\mathrm{Br}(J/\psi\to \gamma \eta_1^{(l,h)}\to \gamma\eta\eta')$ is inversely proportional to the ratio of the total widths of $\eta_1^{(l,h)}$. Given our results and the mixing angle derived by a previous lattice study, whether $\eta_1(1855)$ is assigned to be $\eta_1^{(1)}$ or $\eta_1^{(h)}$, the observed branching fraction $J/\psi\to \eta_1(1855)\to \gamma \eta\eta'$ implies a very large coupling of the octet $\eta_1$ to $\eta\eta'$. This should be investigated in future studies. 
\end{abstract}
\maketitle
\section{Introduction}
Gluons and quarks are fundamental degrees of freedom of Quantum Chromodynamics (QCD). It is expected that gluons can also serve as building blocks to form hadrons. In the quark model picture, the hadrons made up of valence quarks and valence gluons are usually called hybrids. The hybrid mesons with $J^{PC}=1^{-+}$ are most intriguing since this quantum number is prohibited for $q\bar{q}$ states of quark model. Up to now there are three experimental candidates for $I^GJ^{PC}=1^-1^{-+}$ light hybrid mesons, namely, $\pi_1(1400)$~\cite{IHEP-Brussels-LosAlamos-AnnecyLAPP:1988iqi}, $\pi_1(1600)$~\cite{E852:1998mbq,COMPASS:2018uzl,JPAC:2018zyd} and $\pi_1(2105)$~\cite{E852:1998mbq} (details can be found in the latest review~\cite{Chen:2022asf} and the references therein), while lattice QCD studies~\cite{Lacock:1996ny,MILC:1997usn,Mei:2002ip,Bernard:2003jd,Hedditch:2005zf,McNeile:2006bz,Dudek:2013yja,Woss:2020ayi} predict that the mass of isovector $1^{-+}$ hybrid meson has a mass around 1.7-2.2 GeV for light quark masses in a range up to the strange quark mass.
Very recently, the BESIII collaboration reported the first observation of a $I^GJ^{PC}=0^+1^{-+}$ structure $\eta_1(1855)$ through the partial wave analysis of the $J/\psi\to \gamma \eta\eta'$ process~\cite{BESIII:2022riz,BESIII:2022iwi}. The resonance parameters of $\eta_1(1855)$ are determined to be $m_{\eta_1}=1855\pm 9_{-1}^{+6}$ MeV and $\Gamma_{\eta_1}=188\pm 18_{-8}^{+3}$ MeV, and the branching fraction $\mathrm{Br}(J/\psi\to \gamma\eta_1(1855)\to \gamma\eta\eta')$ is $(2.70\pm 0.41_{-0.35}^{+0.16})\times 10^{-6}$. There have been several phenomenological studies on the properties of $\eta_1(1855)$ by assuming it to be an isoscalar light hybrid~\cite{Chen:2022qpd,Qiu:2022ktc,Shastry:2022mhk}, a $K\bar{K}_1(1400)$ molecular state~\cite{Dong:2022cuw,Yang:2022rck}, or a tetraquark state~\cite{Chen:2008ne,Wan:2022xkx}. As far as the hybrid assignment is concerned, there should be two isoscalar $1^{-+}$ mesons in the flavor SU(3) nonet, and a $N_f=2+1$ lattice QCD study~\cite{Dudek:2013yja} does observe two states of masses around 2.16 GeV and 2.33 GeV, respectively in the $0^+1^{-+}$ channel (note the light quark mass here corresponds to a pion mass $m_\pi\sim 390$ MeV). It is noticed that  BESIII also reports a $1^{-+}$ state around 2.2 GeV in the same channel with statistical significance $4.4 \sigma$~\cite{BESIII:2022iwi}.

Since $\eta_1(1855)$ is observed in the $J/\psi$ radiative decay, with regard to the possible hybrid assignment, it is desirable to know the production property of the $1^{-+}$ hybrid meson (named as $\eta_1$ also) in this process, which will provide important information to the nature of $\eta_1(1855)$. This can be investigated in the lattice QCD formalism through the approach similar to the cases of $q\bar{q}$ mesons~\cite{Jiang:2022gnd} and glueballs~\cite{Gui:2012gx,Yang:2013xba,Gui:2019dtm} in $J/\psi$ radiative decays. The key task is to extract the related electromagnetic multipole form factors from the corresponding three-point functions with a vector current insertion, which involve obviously the annihilation diagrams of the light $u,d$ quarks. Therefore, we adopt the distillation method~\cite{Peardon:2009gh} in the practical calculation, which provides a sophisticated scheme for the operator construction and the computation of all-to-all quark propagators. 

This paper is organized as follows: Section~\ref{sec:numerical} presents the details of the numerical calculations of three-point functions, the extraction of the form factors and the interpolation of the form factors to on-shell ones. The discussion of the phenomenological implications of our results can be found in Sec.~\ref{sec:discussion}. Section~\ref{sec:summary} is a brief summary.


\section{Numerical details}\label{sec:numerical}
A large statistics is mandatory for the study of $J/\psi$ radiative decay into light hadrons. Our gauge ensemble of $N_f=2$ degenerate $u,d$ quarks includes 6991 gauge configurations, which are generated on an $L^3\times T=16^3\times 128$ anisotropic lattice with the anisotropy parameter $\xi=a_s/a_t=5.3$ ($a_s$ and $a_t$ are the spatial and temporal lattice spacing, respectively)~\cite{Jiang:2022ffl}. The sea quark mass is tuned to give the pion mass $m_\pi\approx 350$ MeV. The parameters of the gauge ensemble are collected in Table~\ref{tab:config}. For the valence charm quark, we adopt the clover fermion action in Ref.~\cite{CLQCD:2009nvn} and the charm quark mass parameter is set by $(m_{\eta_c}+3m_{J/\psi})/4=3069$ MeV. For each source time slice $\tau\in [0,T-1]$ on each gauge configuration, the perambulators of light $u,d$ quarks are calculated in the Laplacian Heaviside subspace spanned by $N_{\mathrm{vec}}=70$ eigenvectors with lowest eigenvalues.


\begin{table}[t]
    \renewcommand\arraystretch{1.5}
    \caption{Parameters of the gauge ensemble.}
    \label{tab:config}
    \begin{ruledtabular}
        \begin{tabular}{lllllc}
            $L^3\times T$     & $\beta$ & $a_t^{-1}$(GeV) & $\xi$      & $m_\pi$(MeV) & $N_\mathrm{cfg}$ \\\hline
            $16^3 \times 128$ & 2.0     & $6.894(51)$     & $\sim 5.3$ & $348.5(1.0)$ & $6991$           \\
        \end{tabular}
    \end{ruledtabular}
\end{table}

\subsection{Three-point functions}

The partial decay width of $J/\psi\to \gamma \eta_1$ is governed by the on-shell electromagnetic form factors $M_1(Q^2=0)$ and $E_2(Q^2=0)$
($Q^2=-p_\gamma^2$), namely, 
\begin{equation}\label{eq:width}
    \Gamma(J/\psi\to \gamma \eta_1)=\frac{4\alpha}{27}\frac{|\vec{p}_\gamma|}{m_{J/\psi}^2} \left(|M_1(0)|^2+|E_2(0)|^2\right),
\end{equation}
where $\alpha=1/134$ is the fine structure constant at the charm quark mass scale, $\vec{p}_\gamma$ is the momentum of the final state photon with $|\vec{p}_\gamma|=(m^2_{J/\psi}-m_{\eta_1}^2)/(2m_{J/\psi})$ in the rest frame of $J/\psi$. These on-shell form factors can be obtained by the $Q^2\to 0$ interpolation or extrapolation of the form factors $M_1(Q^2)$ and $E_2(Q^2)$, which are defined through the multipole decomposition of the transition matrix elements $\langle \eta_1(p',\lambda')|j_\mathrm{em}^\mu(0)|J/\psi(p,\lambda)\rangle$ (see appendix \ref{sec:form-factors} and also Ref.~\cite{Dudek:2006ej,Dudek:2009kk}).
These matrix elements can be extracted from the following three-point functions 
\begin{eqnarray}\label{eq:three-point}
   &&\Gamma^{(3)}_{i\mu j}(\vec{p},\vec{p}',t,t')=\sum\limits_{\vec{x}} e^{-i\vec{q}\cdot\vec{x}}\nonumber\\
   &\times & \langle\Omega|T\mathcal{O}_{\eta_1}^i(\vec{p'},t)j^\mu_\mathrm{em}(\vec{x},t')\mathcal{O}_{J/\psi}^{j\dagger}(\vec{p},0)|\Omega\rangle, 
\end{eqnarray} 
where $j^\mu_\mathrm{em}$ is the electromagnetic current of quarks, $\mathcal{O}_{\eta_1}^i(\vec{p},t)$ and $\mathcal{O}_{J/\psi}^j(\vec{p},t)$ are the interpolation operators generating $\eta_1$ and $J/\psi$ states with a spatial momentum $\vec{p}$. Therefore, the major numerical task is to calculate these three-point functions from lattice QCD. 

Our lattice setup has the exact SU(2) isospin symmetry. The lattice operator for the isoscalar $\eta_1$ takes the form $\mathcal{O}_{\eta_1}^i=\frac{1}{\sqrt{2}}\epsilon^{ijk}\left(\bar{u}\gamma_j{B}_k u+\bar{d}\gamma_j{B}_k d \right)$, where the chromomagnetic field strength $B_k$ is constructed by the proper combination of the gauge covariant spatial derivatives on the lattice~\cite{Dudek:2013yja}. For the operator $\mathcal{O}_{J/\psi}^i$ we use the conventional $\bar{c}\gamma^i c$-type operator. 
In order to avoid the complication that the momentum projected operator $\mathcal{O}_{\eta_1}^i$ can couple to states with quantum numbers other than $1^{-+}$~\cite{Thomas:2011rh}, the three-point functions 
in Eq.~(\ref{eq:three-point}) are calculated practically in the rest frame of $\eta_1$ with $J/\psi$ moving at different spatial momenta $\vec{p}$.
It has been tested that the dispersion relation of $J/\psi$ satisfies the continuum form very well for all the $\vec{p}$ modes involved~\cite{Jiang:2022ffl}. 

We only consider the initial state radiation and ignore the case that the photon is emitted from quarks in the final state, so the electromagnetic current $j^\mu_\mathrm{em}$ involves charm quarks, namely, $j^\mu_\mathrm{em}=Z_V \bar{c}\gamma^\mu c$ (the electric charge of the charm quark $Q_c=\frac{2}{3}e$ has been absorbed in the prefactor in Eq.~(\ref{eq:width})). Here $Z_V$ is the renormalization 
constant of the current, since $j^\mu_\mathrm{em}$ is not a conserved vector current operator on the lattice. In practice, only the spatial components of $j^\mu_\mathrm{em}$ is involved, and its renormalization constant $Z_V^s=1.118(4)$~\cite{Jiang:2022gnd} is incorporated implicitly into the expressions in the rest part of this work.

\begin{figure}[t]
	\includegraphics[height=3cm]{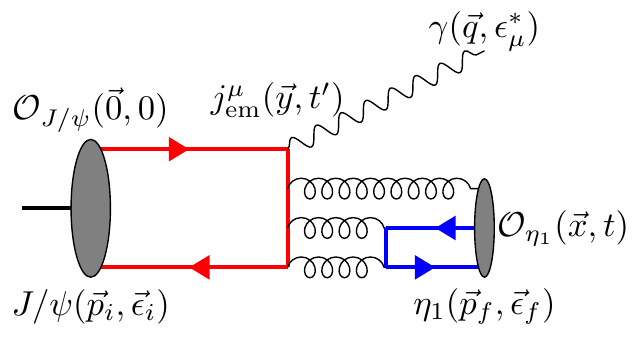}
	\caption{\label{fig:quark-diagram}  The schematic diagram of the process $J/\psi\to \gamma \eta_1$.}
\end{figure}

Figure~\ref{fig:quark-diagram} illustrates the schematic diagram of $\Gamma^{(3)}_{i\mu j}$ after Wick's contraction. It has two separated quark loops, which are actually connected by gluons. The light quark loop on the right hand side can be calculated in the framework of the distillation method. The left part comes from the product of $\mathcal{O}_{J/\psi}$ and the current $j^\mu_\mathrm{em}$, namely,
\begin{equation}
G_{\mu i}(\vec{p},\vec{q};t'+\tau,\tau)=\sum\limits_{\vec{x}} e^{-i\vec{q}\cdot\vec{x}}j_\mathrm{em}^\mu(\vec{x},t'+\tau)O_{J/\psi}^{i\dagger}(\vec{p},\tau).
\end{equation}
which looks very similar to a conventional two-point function of $J/\psi$ and can be calculated independently on each gauge configuration. However, in order for $\Gamma^{(3)}_{i\mu j}$ to have good enough signals, the calculation of $G_{\mu i}$ is highly nontrivial. The conventional momentum source technique turns out to be unfeasible here, because the resulted three-point functions 
\begin{equation}\label{eq:three-point2}
\Gamma_{i\mu j}^{(3)} (\vec{p},\vec{0};t,t')=\frac{1}{T}\sum\limits_{\tau=0}^{T-1}\langle \mathcal{O}_{\eta_1}^i(\vec{0},t+\tau) G_{\mu j}(\vec{p},\vec{p};t'+\tau,\tau) \rangle
\end{equation}
are too noisy even though we have a large gauge ensemble and average over all the time slices $\tau$.


In order to circumvent this difficulty, we calculate $G_{\mu i}$ in the framework of the distillation method. The distillation method provides a gauge covariant smearing scheme for quark fields, taking the charm quark field $c(x)$ for instance,
   $c^{(s)}(\vec{x},t)=\sum\limits_{\vec{y}} \left[VV^\dagger(t)\right](\vec{x},\vec{y}) c(\vec{y},t)$
where $V(t)$ is the matrix whose columns are eigenvectors of the lattice Laplacian operator $-\nabla^2(t)$ at $t$ (we use $N^{(c)}_{\mathrm{vec}}=50$ vectors for charm quarks). 
Therefore, we use the operator $\mathcal{O}_{J/\psi}^i(\vec{p},t)=\sum\limits_{\vec{y}}e^{-i\vec{p}\cdot\vec{y}}[\bar{c}^{(s)}\gamma_i c^{(s)}](\vec{y},t)$ to calculate $G_{\mu i}$, 
whose explicit expression for source time slice at $\tau=0$ is 
\begin{eqnarray}
    &G_{\mu i}(\vec{p},\vec{q};t,0)&=\sum\limits_{\vec{x}}e^{-i\vec{q}\cdot\vec{x}} \mathrm{Tr}\left\{ \gamma_5[S_c V(0)]^\dagger(\vec{x},t)\gamma_5\gamma^\mu\right.\nonumber\\
    &&\times \left. \left[S_c V(0)\right](\vec{x},t)[V^\dagger(0) D(\vec{p})\gamma_i V(0)] \right\},
\end{eqnarray}
where $S_c=\langle c\bar{c}\rangle_U$ is the all-to-all propagator of charm quark for the gauge configuration $U$ and $D(\vec{p})$ is a $3L^3\times 3L^3$ diagonal matrix with the diagonal matrix elements being $\delta_{ij} e^{i\vec{p}\cdot\vec{y}}$ ($\vec{y}$ labels the column or row indices and $i,j=1,2,3$ refer to the color indices). Here we apply the $\gamma_5$-hermiticity of $S_c$, namely, $S_c=\gamma_5 S_c^\dagger\gamma_5$, which implies $[V^\dagger(0)S_c](\vec{x},t)=\gamma_5 [S_cV(0)]^\dagger(\vec{x},t)\gamma_5$, 
such that what we actually calculate is $S_c V(0)$ by solving the linear equation arrays
\begin{equation}
    M[U;m_c][S_c V(0)]=V(0),
\end{equation}
where $M[U;m_c]$ is the fermion matrix in the lattice action of the charm quark. At the source time slice $\tau=0$, we have to solve the linear equation defined by $M[U;m_c]$ for each Dirac index $\alpha=1,2,3,4$ and each column of $V(0)$. In practice, we repeat the above procedure by letting the source time slice $\tau$ running over all the time range, say, $\tau\in[0,T-1]$, to increase the statistics further. This procedure requires 25,600 inversions of $M[U;m_c]$ on each gauge configuration, apart from the calculation of the perambulators of $u,d$ quarks. This prescription turns out to be crucial for us to obtain good signals of the three point functions, from which we can extract the multipole form factors with an acceptable precision.

\subsection{Extraction of form factors}
When $t\gg t'\gg 0$, the three-point function $\Gamma^{(3)}_{i\mu j}(\vec{p},\vec{0};t,t')$ can be parameterized as 
\begin{eqnarray}\label{eq:spectral}
    \Gamma_{ i\mu j}^{(3)} (\vec{p},\vec{0};t,t')&\approx& \frac{Z_{\eta_1}(\vec 0)Z_{J/\psi}^\ast(\vec{p})}{4 m_{\eta_1}E_{J/\psi}(\vec{p})} e^{-m_{\eta_1}(t-t')}e^{-E_{J/\psi}(\vec{p})t'}\nonumber\\
    && \times \mathcal M^{i\mu j}(\vec p),
\end{eqnarray}
where $Z_{\eta_1}(\vec 0)$ and $Z_{J/\psi}(\vec{p})$ come from the matrix element $\langle \Omega|\mathcal{O}_X^i|X(\vec{p},\lambda)\rangle=Z_X(\vec{p})\epsilon_\lambda^i(\vec{p})$ with $X$ referring $J/\psi$ or $\eta_1$ and $\epsilon_\lambda^\mu(\vec{p})$ being its $\lambda$-th polarization vector (note that $Z_X(\vec{p})$ depends on $|\vec{q}|$ since $\mathcal{O}_X(\vec{p})$ is a smeared operator~\cite{Bali:2016lva}), and $\mathcal M^{i\mu j}(\vec p)$ is the desired matrix element
at $\vec p$,
\begin{equation}
    \mathcal M^{i\mu j}(\vec p)= \sum_{\lambda, \lambda'} \epsilon^i_{\lambda'}(\vec{0})\braket{\eta_1(\vec{0}, \lambda')|j^\mu_{\mathrm{em}}(0)|J/\psi(\vec{q}, \lambda)} \epsilon^{\ast j}_\lambda(\vec{p}),
\end{equation}
which is encoded with the multipole form factors $M_1(Q^2)$, $E_2(Q^2)$ etc.

\begin{figure}
    \centering
    \includegraphics[height = 5cm]{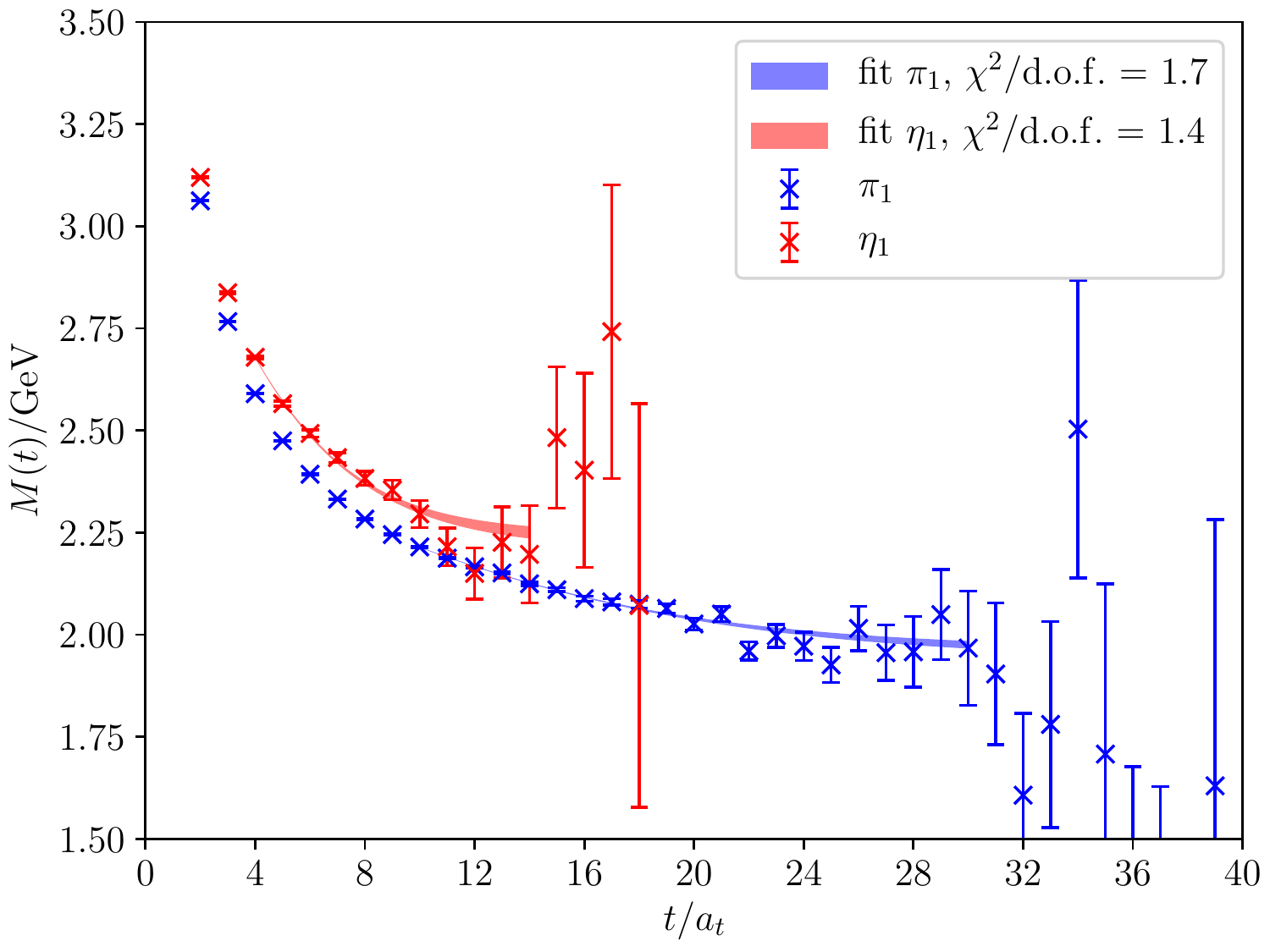}
    \caption{\label{fig:eta1-mass-plot} Effective mass of $\pi_1$ (isovector) and $\eta_1$ (isoscalar), where the fit ranges are $[10, 30]$ and $[4, 14]$ respectively. The shaded curves illustrate the best-fit values with errors using two-mass-term fits.}
    \label{fig:my_label}
\end{figure}

Obviously, in order to extract $\mathcal M^{i\mu j}(\vec p)$, we should know the parameters $Z_X(\vec p), m_{\eta_1}$ and $E_{J/\psi}(\vec p)$, which are actually included in the two-point functions of $\eta_1$ and $J/\psi$, namely, 
\begin{eqnarray}
   && \Gamma^{(2)}_{X}(\vec p, t) = \frac{1}{3T}\sum_{\tau=0}^{T-1}\sum\limits_{i=1}^{3}\braket{\mathcal O_{X, i}(\vec p, \tau+t)\mathcal O_{X, i}^\dagger(\vec p, \tau)}\nonumber\\
   &&= \left(1 + \frac{|\vec p|^2}{3m_X^2}\right)\sum_n\frac{|Z_X^{(n)}(\vec p)|^2}{2E_X^{(n)}(\vec p)} e^{-E_X^{(n)}(\vec p)t},
\end{eqnarray}
where $X$ stands for $J/\psi$ or $\eta_1$ and the source time slice $\tau$ is averaged to increase the statistics. The operators $\mathcal{O}_X$ must be the same as those in the three-point functions $\Gamma_{i\mu j}^{(3)}$, therefore $\Gamma^{(2)}_{X}(\vec p,t)$'s are calculated with the distillation method as well. Since $\eta_1$ is set to be at rest, we only calculate $\Gamma^{(2)}_{\eta_1}(\vec p,t)$ at $\vec p=0$. The effective mass plot is shown in Fig.~\ref{fig:eta1-mass-plot}, where the effective mass of the isovector $1^{-+}$ hybrid state (usually named $\pi_1$) is also plotted for comparison. The effective mass of $\eta_1$ has a much worse signal than that of $\pi_1$ due to the inclusion of disconnected diagrams. Through two-mass-term fits in the time range $t\in [4,14]$ for $\eta_1$ and $t\in [10,30]$ for $\pi_1$, the masses are determined to be $m_{\pi_1} = 1.950(28)\mathrm{GeV}$ and $m_{\eta_1} = 2.230(39)\mathrm{GeV}$, respectively. These results are consistent with those in Ref.~\cite{Dudek:2013yja}.

\begin{figure}[t]
	\includegraphics[height=5cm]{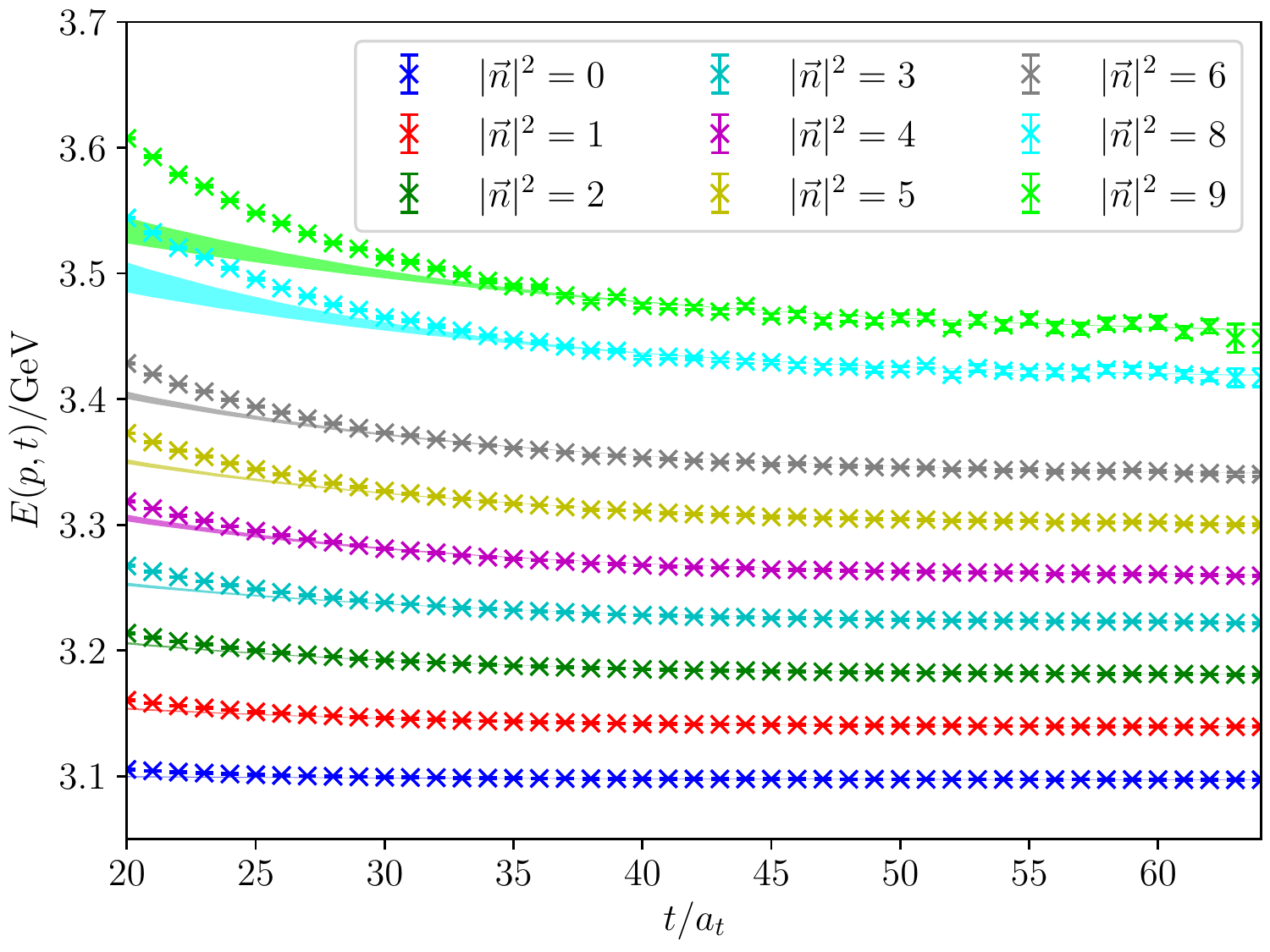}
	\caption{\label{fig:Jpsienergy}  The effective energies of $J/\psi$ at different spatial modes with $|\vec{n}|^2 \le 9$. The data points are the numerical results from $\Gamma^{(2)}_{J/\psi}(\vec{p}, t)$, and the shaded curves illustrate the best-fit values with errors.}
\end{figure}

In order for $Q^2$ to cover the range around $Q^2=0$, the spatial momentum $\vec{p}=\frac{2\pi}{La_s}\vec{n}$ of $J/\psi$ is set to run through all possible modes with $|\vec{n}|^2 \leq 9$ based on the $m_{\eta_1}$ obtained above. Since $\mathcal{O}_{J/\psi}(\vec p,t)$ involved in $\Gamma_{i\mu j}^{(3)}$ is a smeared operator with $N_{\mathrm{vec}}^{(c)}=50$, we also generate the perambulators of the valence charm quark with the same $N_{\mathrm{vec}}^{(c)}$ to calculate $\Gamma^{(2)}_{J/\psi}(\vec p,t)$. The energies $E_{J/\psi}(\vec p) \equiv E_{J/\psi}^{(0)}(\vec p)$ of $J/\psi$ for all the momentum modes involved can be precisely extracted from $\Gamma^{(2)}_{J/\psi}(\vec{p}, t)$ through two-mass-term fits. Fig.~\ref{fig:Jpsienergy} shows the effective energies $E(\vec{p}, t)$ (data points) and the fits (colored bands) at different momentum modes $\vec{n}$ up to $|\vec{n}|^2=9$.   

\begin{figure*}[t]
    \centering
        \includegraphics{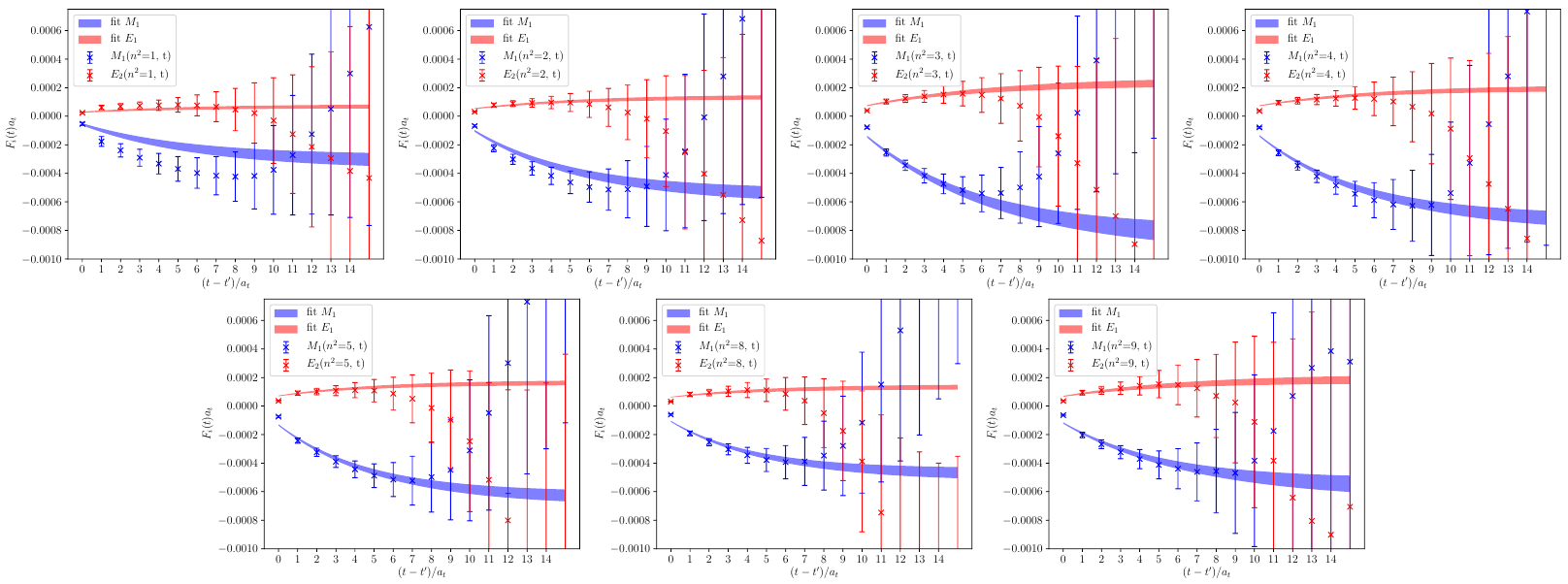}    
    \caption{Multipoles extracted using Eq.~\eqref{eq:sp:form_factor_time_fitter} with momentum modes $|\vec{n}|^2 = 1,2,3,4,5,8,9$, the shaded curves show the fit ranges and best-fit results.}
    \label{fig:sp:formfactor_with_times}
\end{figure*}
Along with the calculated two-point functions of $J/\psi$ and $\eta_1$, the matrix element $\mathcal M^{i\mu j}(\vec p)$ is extracted from the ratio function 
\begin{equation}\label{eq:form_factor_extractor}
   \mathcal M^{i\mu j}(\vec p, t, t') = \frac{\left(1+ \frac{|\vec p|^2}{3m_{J/\psi}^2}\right)Z_{J/\psi}(\vec p) Z_{\eta_1}\Gamma^{(3)}_{i\mu j}(\vec{p}, \vec{0}; t, t')}{\Gamma^{(2)}_{\eta_1}(\vec 0, t - t')\Gamma^{(2)}_{J/\psi}(\vec p, t')}
\end{equation}
which suppresses the contamination from higher states and should be independent of $t$ and $t'$ when ground states dominate. We then make a weighted average value of the function on $t'$ to get larger statistics, and take a convention $\Delta t=t-t'$.
\begin{equation}
   \mathcal M^{i\mu j}(\vec p, \Delta t) = \frac{\sum_{t'=20}^{40}\left(\frac{1}{\Delta_\mathcal{M}^{i\mu j}}\right)^2\mathcal{M}^{i\mu j}(\vec{p},t' + \Delta t,t')}{\sum_{t'=20}^{40}\left(\frac{1}{\Delta_\mathcal{M}^{i\mu j}}\right)^2},
\end{equation}
where $\Delta_\mathcal{M}^{i\mu j}$ is the error of the corresponding ratio function, and the weight is $\left(\frac{1}{\Delta_\mathcal{M}^{i\mu j}}\right)^2$ to make the average value equal to the least square fit result using a constant. $t'\in[20,40]$ indicates the ``fitting window'' in this step. Subsequently, We can extract the form factors $M_1(Q^2,\Delta t)$ and $E_2(Q^2,\Delta t)$ from the linear combination of matrix elements $\mathcal{M}^{i\mu j}(Q^2,\Delta t)$ with specific values of $i$, $\mu$ and $j$. Thus, we can get a similar parameterization for form factors
\begin{equation}\label{eq:sp:form_factor_time_fitter}
    F_i(Q^2,\Delta t)\approx F_i(Q^2)+e^{-\delta m\Delta t},
\end{equation}
where $F_i$ refers to $M_1$ or $E_2$. Note that $Q^2$ is related to $\vec{p}$ by $Q^2=2 m_{\eta_1} E_{J/\psi}(\vec p) -m_{J/\psi}^2-m_{\eta_1}^2$ here. Since $\Gamma^{(3)}_{i\mu j}(\vec{p},\vec{0}; t, t')$ is contributed totally by the disconnected quark diagrams, the signal of $\mathcal{M}^{i\mu j}(\vec p, t'+\Delta t, t')$ becomes very noisy when $\Delta t\gtrsim 10$ and before a clear plateau appears. Therefore, the resulted $M_1(Q^2, \Delta t)$ and $E_2(Q^2, \Delta t)$ have residual time dependence which is absorbed in an additional exponential term in \eqref{eq:sp:form_factor_time_fitter}. We use this equation as the fitting formula to obtain the value of $F_i$ and the corresponding error is acquired from jackknife resampling. Fig.~\ref{fig:sp:formfactor_with_times} shows the $\Delta t$ dependency of $M_1(Q^2, \Delta t)$ and $E_2(Q^2, \Delta t)$, whose $\delta m$ values are listed in Table \ref{tab:delta-m-fits}.

\begin{table}
    \centering
    \begin{ruledtabular}
        \begin{tabular}{ccccc}
            $|n|^2$ & $\delta m$/GeV & $M_1$/GeV & $E_2$/GeV & $\chi^2 / \mathrm{d.o.f.}$ \\
            \hline
            $1$ & $1.19(33)$ & $-2.22(73)$ & $0.49(20)$ & $0.6$\\
            $2$ & $1.25(23)$ & $-3.89(74)$ & $0.95(23)$ & $1.0$\\
            $3$ & $1.08(25)$ & $-6.0(1.3)$ & $1.69(45)$ & $1.0$\\
            $4$ & $1.21(22)$ & $-5.21(87)$ & $1.38(29)$ & $1.1$\\
            $5$ & $1.39(24)$ & $-4.49(69)$ & $1.14(24)$ & $1.2$\\
            $8$ & $1.46(33)$ & $-3.32(63)$ & $0.93(25)$ & $1.0$\\
            $9$ & $1.16(38)$ & $-4.0(1.1)$ & $1.33(44)$ & $1.0$
        \end{tabular}
    \end{ruledtabular}
    \caption{Fitted values of the form factors $M_1(Q^2)$, $E_2(Q^2)$ and $\delta m$. The values of $\chi^2/\mathrm{d.o.f.}$ at different $Q^2$ are also given.}
    \label{tab:delta-m-fits}
\end{table}
The fitted parameters, such as $M_1(Q^2)$, $E_2(Q^2)$ and $\delta m$ are listed in Table \ref{tab:delta-m-fits}, where one can see that the values of $\delta m$ at different $Q^2$ ($\vec{n}^2$) are more or less the same value around 1.2-1.3 GeV. This seems a reasonable value. There are quenched lattice QCD calculations of the masses of the first excited $1^{-+}$ strangeonium-like~\cite{Ma:2020bex} and charmonium-like states~\cite{Ma:2019hsm}, which show that the mass differences of the first excited hybrid states and the ground state hybrids are roughly 1.2-1.3 GeV.

\subsection{On-shell form factors and partial decay width}
\begin{figure}[t!]
	\includegraphics[height=5cm]{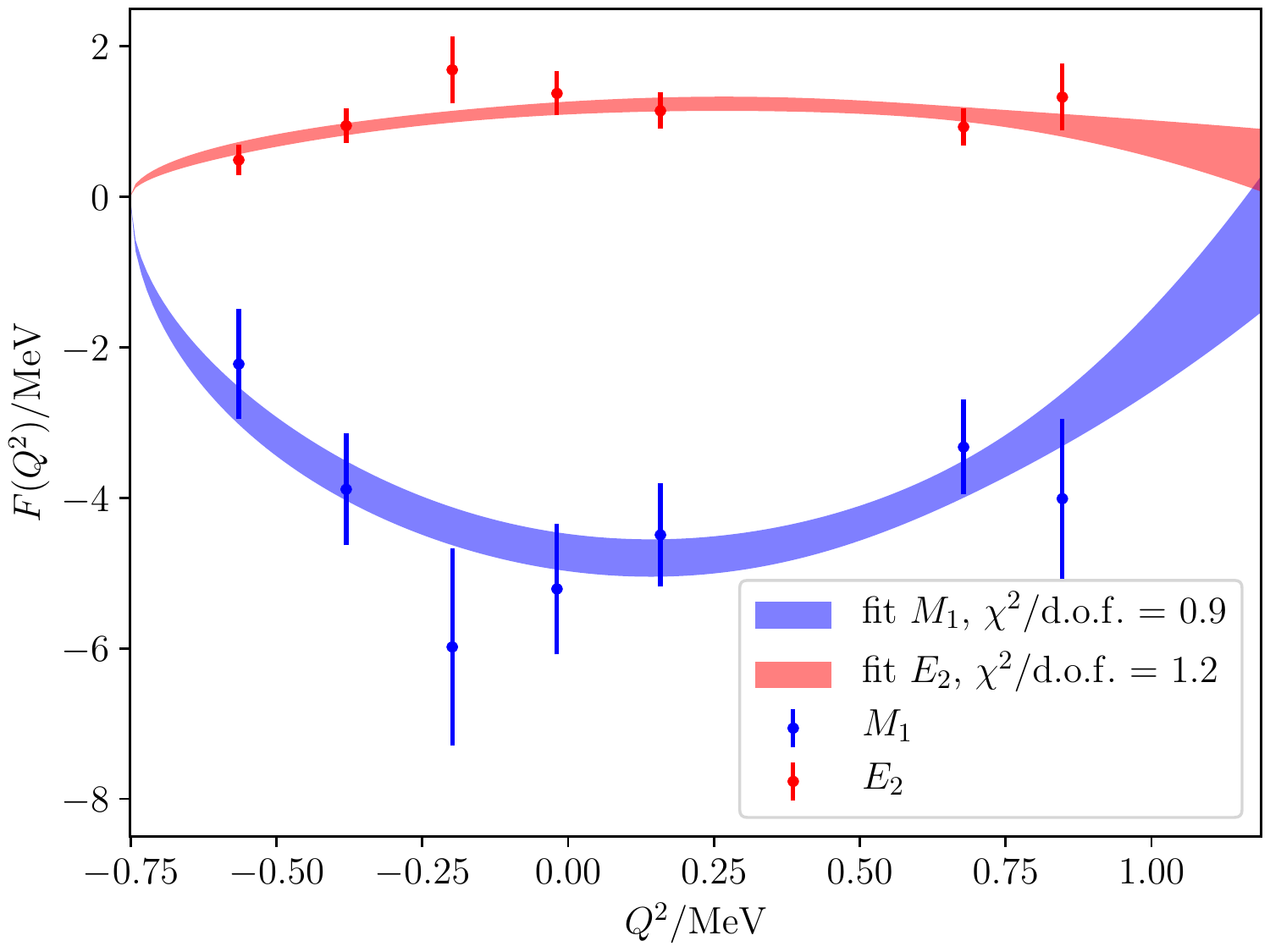}
	\caption{\label{fig:formfactor} Form factors $M_1(Q^2)$ and $E_2(Q^2)$ are plotted with respect to $Q^2$ as data points. The shaded curve illustrates the interpolation using Eq.~(\ref{eq:form_factor_fitter}).}
\end{figure}

After they are determined at different values of $Q^2$, $M_1(Q^2)$ and $E_2(Q^2)$ should be interpolated to the on-shell values at $Q^2=0$, which are required to predict the partial decay width using Eq.~(\ref{eq:width}). If a new Lorentz invariant variable 
\begin{eqnarray}
    \Omega &=& (p\cdot p')^2 - m^2 m'^2 \nonumber\\
           &=& \frac14 [(m + m')^2 + Q^2][(m - m')^2 + Q^2],\label{eq:omega-fn}
\end{eqnarray}
is introduced, one can shown that $M_1(Q^2)$ and $E_2(Q^2)$ are proportional to $\sqrt{\Omega}$, namely,
\begin{eqnarray}\label{eq:sp:form-factor}
    M_1(Q^2) & = & -\frac1{\sqrt{2}}\frac{\sqrt\Omega}{mm'}\left(m G_1(Q^2) + m' G_2(Q^2)\right),\nonumber\\
    E_2(Q^2) & = & ~\frac{1}{\sqrt{2}}\frac{\sqrt\Omega}{mm'}\left(m G_1(Q^2) - m' G_2(Q^2)\right),
\end{eqnarray}
where the form factors $G_1(Q^2)$ and $G_2(Q^2)$ are defined in Eq.~(\ref{eq:sp:form_factor_expand}) (for details see Appendix \ref{sec:form-factors}). Obviously, $M_1(Q^2)$ and $E_2(Q^2)$ go to zero when $\sqrt{\Omega}\to 0$. This provides an additional constraint for the $Q^2$-interpolation. When putting $m=m_{J/\psi}$ and $m'=m_{\eta_1}$ back to the above expressions, we have $\sqrt{\Omega(Q^2)}=m_{\eta_1}|\vec{q}|$ in the rest frame of $\eta_1$. Therefore, it is convenient to introduce a dimensionless function of $Q^2$
\begin{equation}
  v(Q^2) \equiv \sqrt{\Omega(Q^2)}/(m_{J/\psi}m_{\eta_1})=|\vec{q}|/m_{J/\psi}, 
\end{equation}
whose maximum value is $v_\mathrm{max}(Q^2)\approx 0.49$ for the momentum $\vec{q}$ involved in this study. Note that the form factors $G_i(Q^2)$ have no singularities when $Q^2>-m_{J/\psi}^2$. They can be expressed as polynomials of $Q^2$, and certainly polynomials of $v^2(Q^2)$  
\begin{equation}
    G_i(Q^2) = \alpha_i + \beta_i v^2(Q^2) + \delta_i v^4(Q^2) + \mathcal O(v^6(Q^2)),
\end{equation}
where the terms up to $\mathcal{O}(v^4(Q^2))$ are kept, since our kinematic configuration that $\eta_1$ is at rest and $J/\psi$ moves with a momentum $\vec{q}$, we have a dimensionless quantity, the velocity of $J/\psi$, $v(Q^2)=\sqrt{\Omega}/(m_{\eta_1}m_{J/\psi})=|\vec{q}|/m_{J/\psi}<0.494$ for the values of $\vec{q}$ involved, $v_\mathrm{max}^6(Q^2)\sim 1.4\%$ is already much smaller than our statistical errors. Finally, using Eq.~(\ref{eq:sp:form-factor}) we have the interpolation functions for $M_1$ and $E_2$
\begin{equation}\label{eq:form_factor_fitter}
    F_i(Q^2) = v(Q^2)(a_i + b_i v^2(Q^2) + c_i v^4(Q^2))
\end{equation}
with the constraint $F_i(Q^2)=0$ at $v(Q^2)=|\vec{q}|/m_{J/\psi}=0$ in the rest frame of $\eta_1$, as suggested by Ref.~\cite{Yang:2012mya}. Figure~\ref{fig:formfactor} shows the $Q^2$ dependence of $M_1(Q^2)$ and $E_2(Q^2)$ in the region where we are working. 
The interpolation using Eq.~(\ref{eq:form_factor_fitter}) is also illustrated as shaded band in Fig.~\ref{fig:formfactor} with the best-fit parameters (the width of the band shows the interpolation error). Thus we get
\begin{eqnarray}\label{eq:ff-final}
    M_1(0) &= -4.73(74) \mathrm{MeV}\nonumber\\ 
    E_2(0) &= 1.18(22) \mathrm{MeV}.
\end{eqnarray}
Putting these values into Eq.~(\ref{eq:width}), the partial width is predicted to be 
\begin{equation}\label{eq:width-value}
    \Gamma(J/\psi\to \gamma \eta_1)= 2.04(61)~~\mathrm{eV}
\end{equation}
(using the $\eta_1$ mass $m_{\eta_1} = 2.230(39)~\mathrm{GeV}$). Note that the form factors in Eq.~(\ref{eq:ff-final}) are obtained by assuming $\eta_1$ to be a stable particle, while it must be a resonance in principle. For a resonance $R$ of parameters $(m_R, \Gamma_R)$, a more systematic approach to derive the form factor $f_R(Q^2)$ from lattice QCD has been proposed in Refs.~\cite{Briceno:2015csa,Briceno:2016kkp,Briceno:2017max,Alexandrou:2018jbt,Briceno:2021xlc,Radhakrishnan:2022ubg} where the finite volume correction are thoroughly discussed for the $1+\mathcal{J}\to 2$ type transitions with $\mathcal{J}$ being a local current, especially for the case that a resonance can appear in the final two hadron system. However, this approach is unfeasible yet for the processes $J/\psi\to \gamma+\mathrm{light hadron(s)}$ that take place solely through quark annihilation diagrams, because the low precision of $\Gamma_{i\mu j}^{(3)}$ cannot afford that sophisticated treatment. Fortunately, some examples~\cite{Briceno:2015csa,Briceno:2021xlc,Radhakrishnan:2022ubg} indicate that the finite volume correction to the form factors of a narrow resonance $R$ is $\mathcal{O}(\Gamma_R/m_R)$ when $R$ is treated as a stable particle. If $\Gamma_{\eta_1}/m_{\eta_1}$ in the $N_f=2$ case is similar to or even smaller than that of $\eta_1(1855)$, the form factors in Eq.~(\ref{eq:ff-final}) may be taken as approximations for those of the resonant $\eta_1$ with regard to their large statistical uncertainties of roughly 15\%-20\%.  

\section{Discussion}\label{sec:discussion}
Although obtained for $N_f=2$, the form factors in Eq.~(\ref{eq:ff-final}) can be applied to the discussion of the physical SU(3) case. 
In $J/\psi$ radiative decays, the final state light hadron ($\eta_1$ here) is produced by the gluons from 
the $c\bar{c}$ annihilation, and thereby must be a flavor singlet (isoscalar for $N_f=2$ and SU(3) singlet for $N_f=3$). 
If the flavor wave function of the light hadron 
is properly normalized, the underlying gluonic dynamics is usually independent of $N_f$ except for the $U_A(1)$ anomaly 
relevant interaction. In this sense, the form factors in Eq.~(\ref{eq:ff-final}) can be good approximations of 
the SU(3) flavor singlet $\eta_1^{(1)}$ up to a kinematic factor owing to the mass mismatch (see below). Due to 
the flavor SU(3) breaking, there should be two isoscalar mass eigenstates (denoted by $\eta_1^{(l)}$ for the lighter
one and $\eta_1^{(h)}$ for the heavier one), which are the admixtures of the singlet $\eta_1^{(1)}$ and the $I=0$ octet $\eta_1^{(8)}$
through a mixing angle $\theta$, namely, 
\begin{equation}
\left(\begin{array}{c} |\eta_1^{(l)}\rangle \\ |\eta_1^{(h)}\rangle
\end{array}\right)
=\left(\begin{array}{cc} \cos \theta & -\sin\theta  \\ \sin\theta & \cos\theta
\end{array}\right)
\left(\begin{array}{c} |\eta_1^{(8)}\rangle \\ |\eta_1^{(1)}\rangle
\end{array}\right).
\end{equation}
On the other hand, the masses of $\eta_1^{(l,h)}$ can be different from $m_{\eta_1}$ in this study, we should consider the 
correction factor due to the mass mismatch. According to Eq.~(\ref{eq:sp:form-factor}), one has 
\begin{equation}
    M_1^2(0)+E_2^2(0)=|\vec{q}|^2 \frac{m_{J/\psi}^2}{m_{\eta_1}^2}\left(G_1^2(0)+\frac{m_{\eta_1}^2}{m_{J/\psi}^2}G_2^2(0)\right).
\end{equation}
Since the form factors $G_i(Q^2)$ are functions of $Q^2$ and are regular around $Q^2=0$, it is expected the form factors $G_i(0)$ for $i=1,2$ are insensitive to $m_{\eta_1}$ in the range $m_{\eta_1}\sim 2$ GeV. For the case of this study, $G_1^2(0)$ is a few times larger than $G_2^2(0)$, such that from Eq.~\eqref{eq:width}, the $m_{\eta_1}$ dependence is approximately $\Gamma\propto |\vec{q}|^3/m_{\eta_1}^2$. 
Thus one has the following partial widths,
\begin{eqnarray}\label{eq:hlwidth}
    \Gamma(J/\psi\to \gamma \eta_1^{(l)})&=&\chi^{(l)}\Gamma(J/\psi\to \gamma \eta_1) \sin^2 \theta\nonumber\\
    \Gamma(J/\psi\to \gamma \eta_1^{(h)})&=&\chi^{(h)}\Gamma(J/\psi\to \gamma \eta_1) \cos^2 \theta
\end{eqnarray}
where $\chi^{(x)}=\frac{m_{\eta_1}^2 |\vec{p}_\gamma(\eta_1^{(x)})|^3}{m^2_{\eta_1^{(x)}} |\vec{p}_\gamma(\eta_1)|^3}$ is the compensating kinematic 
factor due to the mass mismatch of $\eta_1$ and $\eta_1^{(x)}$.

As for the $\eta\eta'$ decay mode where $\eta_1(1855)$ is observed, since it must be a flavor octet, the flavor SU(3) symmetry implies the decay $\eta_1^{(x)}\to\eta\eta'$ takes place only through its octet component, namely, the decay amplitudes satisfy 
\begin{eqnarray}\label{eq:coupling}
    \langle \eta\eta'|H_I|\eta_1^{(l)}\rangle&=&\cos\theta\langle\eta\eta'|H_I|\eta^{(8)}\rangle\equiv 2g\cos\theta \vec{\epsilon}\cdot\vec{k}^{(l)}\nonumber\\
    \langle \eta\eta'|H_I|\eta_1^{(h)}\rangle&=&\sin\theta\langle\eta\eta'|H_I|\eta^{(8)}\rangle\equiv 2g\sin\theta \vec{\epsilon}\cdot\vec{k}^{(h)},\nonumber\\
\end{eqnarray}
where $g$ is the effective coupling, $\vec{\epsilon}$ is the polarization vector of $\eta_1^{(x)}$ and $\vec{k}^{(x)}$ is the momentum of $\eta\eta'$ in the $\eta^{(x)}$ decay. 
Thus we obtain the ratio
\begin{equation}\label{eq:ratio}
    r=\frac{\mathrm{Br}(J/\psi\to \gamma\eta_1^{(l)}\to\gamma\eta\eta')}{\mathrm{Br}(J/\psi\to \gamma\eta_1^{(h)}\to\gamma\eta\eta')}=\frac{\chi^{(l)}|\vec{k}^{(l)}|^3m^2_{\eta_1^{(h)}}}{\chi^{(h)}|\vec{k}^{(h)}|^3 m^2_{\eta_1^{(l)}}}
    \frac{\Gamma_{\eta_1^{(h)}}}{\Gamma_{\eta_1^{(l)}}}
\end{equation}
which is free from $\theta$ but depends solely on the masses and widths of $\eta_1^{(l)}$ and $\eta_1^{(h)}$. If the mass difference of $\eta_1^{(l)}$ and $\eta_1^{(h)}$ is not too large, the kinematic factor in the above equation is $\mathcal{O}(1)$, such that one has $r\sim \mathcal{O}(1) \frac{\Gamma_{\eta_1^{(h)}}}{\Gamma_{\eta_1^{(l)}}}$.

The lattice QCD study in Ref.~\cite{Dudek:2013yja} observes $\eta_1^{(l)}$ and $\eta_1^{(h)}$ of masses roughly 2.16 GeV and 2.33 GeV (at $m_\pi\approx 391~\mathrm{MeV}$), respectively.
They can be admixtures of the flavor singlet
$\eta_1^{(1)}$ and the flavor octet $\eta_1^{(8)}$ through a mixing angle $\theta$, 
\begin{equation}
\left(\begin{array}{c} |\eta_1^{(l)}\rangle \\ |\eta_1^{(h)}\rangle
\end{array}\right)
=\left(\begin{array}{cc} \cos \theta & -\sin\theta  \\ \sin\theta & \cos\theta
\end{array}\right)
\left(\begin{array}{c} |\eta_1^{(8)}\rangle \\ |\eta_1^{(1)}\rangle
\end{array}\right),
\end{equation}
or equivalently the admixtures of $s\bar{s}$ and $n\bar{n}=(u\bar{u}+d\bar{d})/\sqrt{2}$ through 
a mixing angle $\alpha$,
\begin{equation}
\left(\begin{array}{c} |\eta_1^{(l)}\rangle \\ |\eta_1^{(h)}\rangle
\end{array}\right)
=\left(\begin{array}{cc} \cos \alpha & -\sin\alpha  \\ \sin\alpha & \cos\alpha
\end{array}\right)
\left(\begin{array}{c} |n\bar{n}\rangle \\ |s\bar{s}\rangle
\end{array}\right).
\end{equation}
If the flavor wave functions of $\eta_1^{(1)}$ and $\eta_1^{(8)}$ are defined as 
\begin{eqnarray}
    |\eta_1^{(1)}\rangle &=&\frac{1}{\sqrt{3}}\left(|u\bar{u}\rangle+|d\bar{d}\rangle+~|s\bar{s}\rangle\right)\nonumber\\
    |\eta_1^{(8)}\rangle &=&\frac{1}{\sqrt{6}}\left(|u\bar{u}\rangle+|d\bar{d}\rangle-2|s\bar{s}\rangle\right),
\end{eqnarray}
one can easily show that $\theta$ is related to $\alpha$ by $\theta=\alpha-54.7^\circ$. This convention for the mixing angle $\alpha$ is the same as that in Ref.~\cite{Dudek:2013yja} where $\alpha$ is determined to be roughly $\alpha=22.7(2.1)^\circ$ (averaged over the values on the three lattices involved), such that one has $\theta\approx -32.0(2.1)^\circ$. This indicates a large mixing of $\eta_1^{(8)}$ and $\eta_1^{(1)}$. Using Eq.~(\ref{eq:hlwidth}), the $J/\psi$ total width $\Gamma_\mathrm{tot}=92.6(1.7)$ keV~\cite{Zyla:2020zbs} and the observed branching fraction $\mathrm{Br}(J/\psi\to \gamma\eta_1(1855)\to \gamma\eta\eta')=(2.70\pm 0.41_{-0.35}^{+0.16})\times 10^{-6}$~\cite{BESIII:2022riz}, we get
\begin{eqnarray}
\Gamma(J/\psi\to \gamma \eta_1(1855)) &=& (2.0\pm 0.7)~\mathrm{eV}\nonumber\\
\mathrm{Br}(J/\psi\to \gamma \eta_1(1855))&=&(2.1\pm 0.7)\times 10^{-5}\nonumber\\
\mathrm{Br}(\eta_1(1855)\to \eta\eta')&=& (13\pm 5)\%
\end{eqnarray}
if $\eta_1(1855)$ is assigned to be $\eta_1^{(l)}$, and 
\begin{eqnarray}
\Gamma(J/\psi\to \gamma \eta_1(1855)) &=& (5.0\pm 1.6)~\mathrm{eV}\nonumber\\
\mathrm{Br}(J/\psi\to \gamma \eta_1(1855))&=&(5.4\pm 1.8)\times 10^{-5}\nonumber\\
\mathrm{Br}(\eta_1(1855)\to \eta\eta')&=& (5.0\pm 1.9)\%
\end{eqnarray}
if $\eta_1(1855)$ is assigned to be $\eta_1^{(h)}$. 

Obviously, the existence of the other $\eta_1$ state (or not) is crucial for the nature of $\eta_1(1855)$ to be unravelled. We notice BESIII also reports a weak ($4.4\sigma$) signal of $1^{-+}$ component around 2.2 GeV~\cite{BESIII:2022iwi}. But its existence need to be confirmed. On the other hand, if $\eta_1(1855)$ is surely a hybrid state (either $\eta_1^{(l)}$ or $\eta_1^{(h)}$), the results and the discussion imply that the octet $\eta_1^{(8)}$ couples strongly to $\eta\eta'$, namely, the effective coupling in Eq.~(\ref{eq:coupling}) is roughly $g = 5.0(1.0)$ (note the effective coupling $g_{\rho\pi\pi}\approx 6.0$ for the decay process $\rho\to \pi\pi$). Although the possible enhancement by QCD $U_A(1)$ anomaly~\cite{Chen:2022qpd,Qiu:2022ktc}, this is really a large coupling and should be understood when comparing with the significantly small coupling of its isovector partner $\pi_1$ to $\eta'\pi$, which is expected by phenomenological studies~\cite{Page:1996rj,Page:1998gz} and estimated by lattice QCD calculations~\cite{McNeile:2006bz,Woss:2020ayi}.

\section{Summary}\label{sec:summary}
Based on a large gauge ensemble of $N_f=2$ dynamical quarks at $m_\pi\approx 350$ MeV, we perform the first theoretical calculation of $\Gamma(J/\psi\to \gamma \eta_1)$ where $\eta_1$ is the light flavor singlet $1^{-+}$ hybrid. The related three-point functions are contributed totally from disconnected quark diagrams, which are dealt with using the distillation method. The on-shell electromagnetic form factors are determined to be $M_1(0)=-4.73(74)$ MeV and $E_2(0)=1.18$ MeV, which give $\Gamma(J/\psi\to \gamma \eta_1)=2.04(61)~\mathrm{eV}$ for $m_{\eta_1}=2.23(4)$ GeV. These results are applicable to discuss the production rates of the two mass eigenstates $\eta_1^{(l)}$ and $\eta_1^{(h)}$ in the SU(3) case, if the singlet-octet mixing angle is known. As for $\eta_1(1855)$ observed by BESIII, its hybrid assignment depends strongly on the existence of its mass partner. It should be emphasized that the ratio of the branching fractions $\mathrm{Br}(J/\psi\to \gamma \eta_1^{(l,h)}\to \gamma\eta\eta')$ is inversely proportional to the ratio of the total widths of $\eta_1^{(l,h)}$. This can be used as one of the criteria to identify $\eta^{(l,h)}$ experimentally. If $\eta_1(1855)$ is a hybrid for sure, our results and the mixing angle $\theta$ determined in Ref.~\cite{Dudek:2013yja} indicate that the coupling of the octet $1^{-+}$ hybrid $\eta_1^{(8)}$ to $\eta\eta'$ is very large. This is interesting and worthy of an investigation in depth. Throughout our calculation, $\eta_1$ is tentatively viewed as a stable particle. This surely introduce theoretical uncertainties which cannot be accessed in the present stage, but should be explored in future works. Nevertheless, this study provides the first valuable theoretical predictions for this intriguing topic from lattice QCD.

\section{Acknowledgement}
We thank Qiang Zhao for valuable discussions. This work is supported by the National Key Research and Development Program of China (No. 2020YFA0406400), the Strategic Priority Research Program of Chinese Academy of Sciences (No. XDB34030302) and the National Natural Science Foundation of China (NNSFC) under Grants No.11935017, No.12075253, No.12070131001 (CRC 110 by DFG and NNSFC), No.12175063, No.12205311 and No.12293065. The Chroma software system~\cite{Edwards:2004sx} and QUDA library~\cite{Clark:2009wm,Babich:2011np} are acknowledged. The computations were performed on the HPC clusters at Institute of High Energy Physics (Beijing) and China Spallation Neutron Source (Dongguan), and the ORISE computing environment.

\bibliography{ref}

\begin{thebibliography}{48}%
\makeatletter
\providecommand \@ifxundefined [1]{%
 \@ifx{#1\undefined}
}%
\providecommand \@ifnum [1]{%
 \ifnum #1\expandafter \@firstoftwo
 \else \expandafter \@secondoftwo
 \fi
}%
\providecommand \@ifx [1]{%
 \ifx #1\expandafter \@firstoftwo
 \else \expandafter \@secondoftwo
 \fi
}%
\providecommand \natexlab [1]{#1}%
\providecommand \enquote  [1]{``#1''}%
\providecommand \bibnamefont  [1]{#1}%
\providecommand \bibfnamefont [1]{#1}%
\providecommand \citenamefont [1]{#1}%
\providecommand \href@noop [0]{\@secondoftwo}%
\providecommand \href [0]{\begingroup \@sanitize@url \@href}%
\providecommand \@href[1]{\@@startlink{#1}\@@href}%
\providecommand \@@href[1]{\endgroup#1\@@endlink}%
\providecommand \@sanitize@url [0]{\catcode `\\12\catcode `\$12\catcode
  `\&12\catcode `\#12\catcode `\^12\catcode `\_12\catcode `\%12\relax}%
\providecommand \@@startlink[1]{}%
\providecommand \@@endlink[0]{}%
\providecommand \url  [0]{\begingroup\@sanitize@url \@url }%
\providecommand \@url [1]{\endgroup\@href {#1}{\urlprefix }}%
\providecommand \urlprefix  [0]{URL }%
\providecommand \Eprint [0]{\href }%
\providecommand \doibase [0]{http://dx.doi.org/}%
\providecommand \selectlanguage [0]{\@gobble}%
\providecommand \bibinfo  [0]{\@secondoftwo}%
\providecommand \bibfield  [0]{\@secondoftwo}%
\providecommand \translation [1]{[#1]}%
\providecommand \BibitemOpen [0]{}%
\providecommand \bibitemStop [0]{}%
\providecommand \bibitemNoStop [0]{.\EOS\space}%
\providecommand \EOS [0]{\spacefactor3000\relax}%
\providecommand \BibitemShut  [1]{\csname bibitem#1\endcsname}%
\let\auto@bib@innerbib\@empty
\bibitem [{\citenamefont {Alde}\ \emph {et~al.}(1988)\citenamefont {Alde} \emph
  {et~al.}}]{IHEP-Brussels-LosAlamos-AnnecyLAPP:1988iqi}%
  \BibitemOpen
  \bibfield  {author} {\bibinfo {author} {\bibfnamefont {D.}~\bibnamefont
  {Alde}} \emph {et~al.} (\bibinfo {collaboration} {IHEP-Brussels-Los
  Alamos-Annecy(LAPP)}),\ }\bibfield  {title} {\enquote {\bibinfo {title}
  {{Evidence for a $1^{-+}$ Exotic Meson}},}\ }\href {\doibase
  10.1016/0370-2693(88)91686-3} {\bibfield  {journal} {\bibinfo  {journal}
  {Phys. Lett. B}\ }\textbf {\bibinfo {volume} {205}},\ \bibinfo {pages} {397}
  (\bibinfo {year} {1988})}\BibitemShut {NoStop}%
\bibitem [{\citenamefont {Adams}\ \emph {et~al.}(1998)\citenamefont {Adams}
  \emph {et~al.}}]{E852:1998mbq}%
  \BibitemOpen
  \bibfield  {author} {\bibinfo {author} {\bibfnamefont {G.~S.}\ \bibnamefont
  {Adams}} \emph {et~al.} (\bibinfo {collaboration} {E852}),\ }\bibfield
  {title} {\enquote {\bibinfo {title} {{Observation of a New $J^{PC} = 1^{-+}$
  Exotic State in the Reaction $\pi^- p \to \pi^+ \pi^- \pi^- p$ at
  18-GeV/$c$}},}\ }\href {\doibase 10.1103/PhysRevLett.81.5760} {\bibfield
  {journal} {\bibinfo  {journal} {Phys. Rev. Lett.}\ }\textbf {\bibinfo
  {volume} {81}},\ \bibinfo {pages} {5760--5763} (\bibinfo {year}
  {1998})}\BibitemShut {NoStop}%
\bibitem [{\citenamefont {Aghasyan}\ \emph {et~al.}(2018)\citenamefont
  {Aghasyan} \emph {et~al.}}]{COMPASS:2018uzl}%
  \BibitemOpen
  \bibfield  {author} {\bibinfo {author} {\bibfnamefont {M.}~\bibnamefont
  {Aghasyan}} \emph {et~al.} (\bibinfo {collaboration} {COMPASS}),\ }\bibfield
  {title} {\enquote {\bibinfo {title} {{Light isovector resonances in $\pi^- p
  \to \pi^-\pi^-\pi^+ p$ at 190 GeV/${\it c}$}},}\ }\href {\doibase
  10.1103/PhysRevD.98.092003} {\bibfield  {journal} {\bibinfo  {journal} {Phys.
  Rev. D}\ }\textbf {\bibinfo {volume} {98}},\ \bibinfo {pages} {092003}
  (\bibinfo {year} {2018})},\ \Eprint {http://arxiv.org/abs/1802.05913}
  {arXiv:1802.05913 [hep-ex]} \BibitemShut {NoStop}%
\bibitem [{\citenamefont {Rodas}\ \emph {et~al.}(2019)\citenamefont {Rodas}
  \emph {et~al.}}]{JPAC:2018zyd}%
  \BibitemOpen
  \bibfield  {author} {\bibinfo {author} {\bibfnamefont {A.}~\bibnamefont
  {Rodas}} \emph {et~al.} (\bibinfo {collaboration} {JPAC}),\ }\bibfield
  {title} {\enquote {\bibinfo {title} {{Determination of the pole position of
  the lightest hybrid meson candidate}},}\ }\href {\doibase
  10.1103/PhysRevLett.122.042002} {\bibfield  {journal} {\bibinfo  {journal}
  {Phys. Rev. Lett.}\ }\textbf {\bibinfo {volume} {122}},\ \bibinfo {pages}
  {042002} (\bibinfo {year} {2019})},\ \Eprint
  {http://arxiv.org/abs/1810.04171} {arXiv:1810.04171 [hep-ph]} \BibitemShut
  {NoStop}%
\bibitem [{\citenamefont {Chen}\ \emph
  {et~al.}(2022{\natexlab{a}})\citenamefont {Chen}, \citenamefont {Chen},
  \citenamefont {Liu}, \citenamefont {Liu},\ and\ \citenamefont
  {Zhu}}]{Chen:2022asf}%
  \BibitemOpen
  \bibfield  {author} {\bibinfo {author} {\bibfnamefont {Hua-Xing}\
  \bibnamefont {Chen}}, \bibinfo {author} {\bibfnamefont {Wei}\ \bibnamefont
  {Chen}}, \bibinfo {author} {\bibfnamefont {Xiang}\ \bibnamefont {Liu}},
  \bibinfo {author} {\bibfnamefont {Yan-Rui}\ \bibnamefont {Liu}}, \ and\
  \bibinfo {author} {\bibfnamefont {Shi-Lin}\ \bibnamefont {Zhu}},\ }\bibfield
  {title} {\enquote {\bibinfo {title} {{An updated review of the new hadron
  states}},}\ }\href@noop {} {\  (\bibinfo {year} {2022}{\natexlab{a}})},\
  \Eprint {http://arxiv.org/abs/2204.02649} {arXiv:2204.02649 [hep-ph]}
  \BibitemShut {NoStop}%
\bibitem [{\citenamefont {Lacock}\ \emph {et~al.}(1997)\citenamefont {Lacock},
  \citenamefont {Michael}, \citenamefont {Boyle},\ and\ \citenamefont
  {Rowland}}]{Lacock:1996ny}%
  \BibitemOpen
  \bibfield  {author} {\bibinfo {author} {\bibfnamefont {P.}~\bibnamefont
  {Lacock}}, \bibinfo {author} {\bibfnamefont {Christopher}\ \bibnamefont
  {Michael}}, \bibinfo {author} {\bibfnamefont {P.}~\bibnamefont {Boyle}}, \
  and\ \bibinfo {author} {\bibfnamefont {P.}~\bibnamefont {Rowland}} (\bibinfo
  {collaboration} {UKQCD}),\ }\bibfield  {title} {\enquote {\bibinfo {title}
  {{Hybrid mesons from quenched QCD}},}\ }\href {\doibase
  10.1016/S0370-2693(97)00384-5} {\bibfield  {journal} {\bibinfo  {journal}
  {Phys. Lett. B}\ }\textbf {\bibinfo {volume} {401}},\ \bibinfo {pages}
  {308--312} (\bibinfo {year} {1997})},\ \Eprint
  {http://arxiv.org/abs/hep-lat/9611011} {arXiv:hep-lat/9611011} \BibitemShut
  {NoStop}%
\bibitem [{\citenamefont {Bernard}\ \emph {et~al.}(1997)\citenamefont {Bernard}
  \emph {et~al.}}]{MILC:1997usn}%
  \BibitemOpen
  \bibfield  {author} {\bibinfo {author} {\bibfnamefont {Claude~W.}\
  \bibnamefont {Bernard}} \emph {et~al.} (\bibinfo {collaboration} {MILC}),\
  }\bibfield  {title} {\enquote {\bibinfo {title} {{Exotic mesons in quenched
  lattice QCD}},}\ }\href {\doibase 10.1103/PhysRevD.56.7039} {\bibfield
  {journal} {\bibinfo  {journal} {Phys. Rev. D}\ }\textbf {\bibinfo {volume}
  {56}},\ \bibinfo {pages} {7039--7051} (\bibinfo {year} {1997})},\ \Eprint
  {http://arxiv.org/abs/hep-lat/9707008} {arXiv:hep-lat/9707008} \BibitemShut
  {NoStop}%
\bibitem [{\citenamefont {Mei}\ and\ \citenamefont {Luo}(2003)}]{Mei:2002ip}%
  \BibitemOpen
  \bibfield  {author} {\bibinfo {author} {\bibfnamefont {Zhong-Hao}\
  \bibnamefont {Mei}}\ and\ \bibinfo {author} {\bibfnamefont {Xiang-Qian}\
  \bibnamefont {Luo}},\ }\bibfield  {title} {\enquote {\bibinfo {title}
  {{Exotic mesons from quantum chromodynamics with improved gluon and quark
  actions on the anisotropic lattice}},}\ }\href {\doibase
  10.1142/S0217751X03017038} {\bibfield  {journal} {\bibinfo  {journal} {Int.
  J. Mod. Phys. A}\ }\textbf {\bibinfo {volume} {18}},\ \bibinfo {pages} {5713}
  (\bibinfo {year} {2003})},\ \Eprint {http://arxiv.org/abs/hep-lat/0206012}
  {arXiv:hep-lat/0206012} \BibitemShut {NoStop}%
\bibitem [{\citenamefont {Bernard}\ \emph {et~al.}(2003)\citenamefont
  {Bernard}, \citenamefont {Burch}, \citenamefont {Gregory}, \citenamefont
  {Toussaint}, \citenamefont {DeTar}, \citenamefont {Osborn}, \citenamefont
  {Gottlieb}, \citenamefont {Heller},\ and\ \citenamefont
  {Sugar}}]{Bernard:2003jd}%
  \BibitemOpen
  \bibfield  {author} {\bibinfo {author} {\bibfnamefont {C.}~\bibnamefont
  {Bernard}}, \bibinfo {author} {\bibfnamefont {T.}~\bibnamefont {Burch}},
  \bibinfo {author} {\bibfnamefont {E.~B.}\ \bibnamefont {Gregory}}, \bibinfo
  {author} {\bibfnamefont {D.}~\bibnamefont {Toussaint}}, \bibinfo {author}
  {\bibfnamefont {Carleton~E.}\ \bibnamefont {DeTar}}, \bibinfo {author}
  {\bibfnamefont {J.}~\bibnamefont {Osborn}}, \bibinfo {author} {\bibfnamefont
  {Steven~A.}\ \bibnamefont {Gottlieb}}, \bibinfo {author} {\bibfnamefont
  {U.~M.}\ \bibnamefont {Heller}}, \ and\ \bibinfo {author} {\bibfnamefont
  {R.}~\bibnamefont {Sugar}},\ }\bibfield  {title} {\enquote {\bibinfo {title}
  {{Lattice calculation of $1^{-+}$ hybrid mesons with improved Kogut-Susskind
  fermions}},}\ }\href {\doibase 10.1103/PhysRevD.68.074505} {\bibfield
  {journal} {\bibinfo  {journal} {Phys. Rev. D}\ }\textbf {\bibinfo {volume}
  {68}},\ \bibinfo {pages} {074505} (\bibinfo {year} {2003})},\ \Eprint
  {http://arxiv.org/abs/hep-lat/0301024} {arXiv:hep-lat/0301024} \BibitemShut
  {NoStop}%
\bibitem [{\citenamefont {Hedditch}\ \emph {et~al.}(2005)\citenamefont
  {Hedditch}, \citenamefont {Kamleh}, \citenamefont {Lasscock}, \citenamefont
  {Leinweber}, \citenamefont {Williams},\ and\ \citenamefont
  {Zanotti}}]{Hedditch:2005zf}%
  \BibitemOpen
  \bibfield  {author} {\bibinfo {author} {\bibfnamefont {J.~N.}\ \bibnamefont
  {Hedditch}}, \bibinfo {author} {\bibfnamefont {W.}~\bibnamefont {Kamleh}},
  \bibinfo {author} {\bibfnamefont {B.~G.}\ \bibnamefont {Lasscock}}, \bibinfo
  {author} {\bibfnamefont {D.~B.}\ \bibnamefont {Leinweber}}, \bibinfo {author}
  {\bibfnamefont {A.~G.}\ \bibnamefont {Williams}}, \ and\ \bibinfo {author}
  {\bibfnamefont {J.~M.}\ \bibnamefont {Zanotti}},\ }\bibfield  {title}
  {\enquote {\bibinfo {title} {{$1^{-+}$ exotic meson at light quark
  masses}},}\ }\href {\doibase 10.1103/PhysRevD.72.114507} {\bibfield
  {journal} {\bibinfo  {journal} {Phys. Rev. D}\ }\textbf {\bibinfo {volume}
  {72}},\ \bibinfo {pages} {114507} (\bibinfo {year} {2005})},\ \Eprint
  {http://arxiv.org/abs/hep-lat/0509106} {arXiv:hep-lat/0509106} \BibitemShut
  {NoStop}%
\bibitem [{\citenamefont {McNeile}\ and\ \citenamefont
  {Michael}(2006)}]{McNeile:2006bz}%
  \BibitemOpen
  \bibfield  {author} {\bibinfo {author} {\bibfnamefont {C.}~\bibnamefont
  {McNeile}}\ and\ \bibinfo {author} {\bibfnamefont {Christopher}\ \bibnamefont
  {Michael}} (\bibinfo {collaboration} {UKQCD}),\ }\bibfield  {title} {\enquote
  {\bibinfo {title} {{Decay width of light quark hybrid meson from the
  lattice}},}\ }\href {\doibase 10.1103/PhysRevD.73.074506} {\bibfield
  {journal} {\bibinfo  {journal} {Phys. Rev. D}\ }\textbf {\bibinfo {volume}
  {73}},\ \bibinfo {pages} {074506} (\bibinfo {year} {2006})},\ \Eprint
  {http://arxiv.org/abs/hep-lat/0603007} {arXiv:hep-lat/0603007} \BibitemShut
  {NoStop}%
\bibitem [{\citenamefont {Dudek}\ \emph {et~al.}(2013)\citenamefont {Dudek},
  \citenamefont {Edwards}, \citenamefont {Guo},\ and\ \citenamefont
  {Thomas}}]{Dudek:2013yja}%
  \BibitemOpen
  \bibfield  {author} {\bibinfo {author} {\bibfnamefont {Jozef~J.}\
  \bibnamefont {Dudek}}, \bibinfo {author} {\bibfnamefont {Robert~G.}\
  \bibnamefont {Edwards}}, \bibinfo {author} {\bibfnamefont {Peng}\
  \bibnamefont {Guo}}, \ and\ \bibinfo {author} {\bibfnamefont
  {Christopher~E.}\ \bibnamefont {Thomas}} (\bibinfo {collaboration} {Hadron
  Spectrum}),\ }\bibfield  {title} {\enquote {\bibinfo {title} {{Toward the
  excited isoscalar meson spectrum from lattice QCD}},}\ }\href {\doibase
  10.1103/PhysRevD.88.094505} {\bibfield  {journal} {\bibinfo  {journal} {Phys.
  Rev. D}\ }\textbf {\bibinfo {volume} {88}},\ \bibinfo {pages} {094505}
  (\bibinfo {year} {2013})},\ \Eprint {http://arxiv.org/abs/1309.2608}
  {arXiv:1309.2608 [hep-lat]} \BibitemShut {NoStop}%
\bibitem [{\citenamefont {Woss}\ \emph {et~al.}(2021)\citenamefont {Woss},
  \citenamefont {Dudek}, \citenamefont {Edwards}, \citenamefont {Thomas},\ and\
  \citenamefont {Wilson}}]{Woss:2020ayi}%
  \BibitemOpen
  \bibfield  {author} {\bibinfo {author} {\bibfnamefont {Antoni~J.}\
  \bibnamefont {Woss}}, \bibinfo {author} {\bibfnamefont {Jozef~J.}\
  \bibnamefont {Dudek}}, \bibinfo {author} {\bibfnamefont {Robert~G.}\
  \bibnamefont {Edwards}}, \bibinfo {author} {\bibfnamefont {Christopher~E.}\
  \bibnamefont {Thomas}}, \ and\ \bibinfo {author} {\bibfnamefont {David~J.}\
  \bibnamefont {Wilson}} (\bibinfo {collaboration} {Hadron Spectrum}),\
  }\bibfield  {title} {\enquote {\bibinfo {title} {{Decays of an exotic
  $1^{-+}$ hybrid meson resonance in QCD}},}\ }\href {\doibase
  10.1103/PhysRevD.103.054502} {\bibfield  {journal} {\bibinfo  {journal}
  {Phys. Rev. D}\ }\textbf {\bibinfo {volume} {103}},\ \bibinfo {pages}
  {054502} (\bibinfo {year} {2021})},\ \Eprint
  {http://arxiv.org/abs/2009.10034} {arXiv:2009.10034 [hep-lat]} \BibitemShut
  {NoStop}%
\bibitem [{\citenamefont {Ablikim}\ \emph
  {et~al.}(2022{\natexlab{a}})\citenamefont {Ablikim} \emph
  {et~al.}}]{BESIII:2022riz}%
  \BibitemOpen
  \bibfield  {author} {\bibinfo {author} {\bibfnamefont {M.}~\bibnamefont
  {Ablikim}} \emph {et~al.} (\bibinfo {collaboration} {BESIII}),\ }\bibfield
  {title} {\enquote {\bibinfo {title} {{Observation of an Isoscalar Resonance
  with Exotic $J^{PC}=1^{-+}$ Quantum Numbers in $J/\psi\to \gamma
  \eta\eta'$'}},}\ }\href {\doibase 10.1103/PhysRevLett.129.192002} {\bibfield
  {journal} {\bibinfo  {journal} {Phys. Rev. Lett.}\ }\textbf {\bibinfo
  {volume} {129}},\ \bibinfo {pages} {192002} (\bibinfo {year}
  {2022}{\natexlab{a}})},\ \Eprint {http://arxiv.org/abs/2202.00621}
  {arXiv:2202.00621 [hep-ex]} \BibitemShut {NoStop}%
\bibitem [{\citenamefont {Ablikim}\ \emph
  {et~al.}(2022{\natexlab{b}})\citenamefont {Ablikim} \emph
  {et~al.}}]{BESIII:2022iwi}%
  \BibitemOpen
  \bibfield  {author} {\bibinfo {author} {\bibfnamefont {M.}~\bibnamefont
  {Ablikim}} \emph {et~al.} (\bibinfo {collaboration} {BESIII}),\ }\bibfield
  {title} {\enquote {\bibinfo {title} {{Partial wave analysis of $J/\psi\to
  \gamma \eta\eta'$}},}\ }\href {\doibase 10.1103/PhysRevD.106.072012}
  {\bibfield  {journal} {\bibinfo  {journal} {Phys. Rev. D}\ }\textbf {\bibinfo
  {volume} {106}},\ \bibinfo {pages} {072012} (\bibinfo {year}
  {2022}{\natexlab{b}})},\ \Eprint {http://arxiv.org/abs/2202.00623}
  {arXiv:2202.00623 [hep-ex]} \BibitemShut {NoStop}%
\bibitem [{\citenamefont {Chen}\ \emph
  {et~al.}(2022{\natexlab{b}})\citenamefont {Chen}, \citenamefont {Su},\ and\
  \citenamefont {Zhu}}]{Chen:2022qpd}%
  \BibitemOpen
  \bibfield  {author} {\bibinfo {author} {\bibfnamefont {Hua-Xing}\
  \bibnamefont {Chen}}, \bibinfo {author} {\bibfnamefont {Niu}\ \bibnamefont
  {Su}}, \ and\ \bibinfo {author} {\bibfnamefont {Shi-Lin}\ \bibnamefont
  {Zhu}},\ }\bibfield  {title} {\enquote {\bibinfo {title} {{QCD Axial Anomaly
  Enhances the $\eta\eta'$ Decay of the Hybrid Candidate $\eta_{1}$(1855)}},}\
  }\href {\doibase 10.1088/0256-307X/39/5/051201} {\bibfield  {journal}
  {\bibinfo  {journal} {Chin. Phys. Lett.}\ }\textbf {\bibinfo {volume} {39}},\
  \bibinfo {pages} {051201} (\bibinfo {year} {2022}{\natexlab{b}})},\ \Eprint
  {http://arxiv.org/abs/2202.04918} {arXiv:2202.04918 [hep-ph]} \BibitemShut
  {NoStop}%
\bibitem [{\citenamefont {Qiu}\ and\ \citenamefont {Zhao}(2022)}]{Qiu:2022ktc}%
  \BibitemOpen
  \bibfield  {author} {\bibinfo {author} {\bibfnamefont {Lin}\ \bibnamefont
  {Qiu}}\ and\ \bibinfo {author} {\bibfnamefont {Qiang}\ \bibnamefont {Zhao}},\
  }\bibfield  {title} {\enquote {\bibinfo {title} {{Towards the establishment
  of the light $J^{P (C )}$=$1^{–(+)}$ hybrid nonet}},}\ }\href {\doibase
  10.1088/1674-1137/ac567e} {\bibfield  {journal} {\bibinfo  {journal} {Chin.
  Phys. C}\ }\textbf {\bibinfo {volume} {46}},\ \bibinfo {pages} {051001}
  (\bibinfo {year} {2022})},\ \Eprint {http://arxiv.org/abs/2202.00904}
  {arXiv:2202.00904 [hep-ph]} \BibitemShut {NoStop}%
\bibitem [{\citenamefont {Shastry}\ \emph {et~al.}(2022)\citenamefont
  {Shastry}, \citenamefont {Fischer},\ and\ \citenamefont
  {Giacosa}}]{Shastry:2022mhk}%
  \BibitemOpen
  \bibfield  {author} {\bibinfo {author} {\bibfnamefont {Vanamali}\
  \bibnamefont {Shastry}}, \bibinfo {author} {\bibfnamefont {Christian~S.}\
  \bibnamefont {Fischer}}, \ and\ \bibinfo {author} {\bibfnamefont {Francesco}\
  \bibnamefont {Giacosa}},\ }\bibfield  {title} {\enquote {\bibinfo {title}
  {{The phenomenology of the exotic hybrid nonet with $\pi_1(1600)$ and
  $\eta_1(1855)$}},}\ }\href {\doibase 10.1016/j.physletb.2022.137478}
  {\bibfield  {journal} {\bibinfo  {journal} {Phys. Lett. B}\ }\textbf
  {\bibinfo {volume} {834}},\ \bibinfo {pages} {137478} (\bibinfo {year}
  {2022})},\ \Eprint {http://arxiv.org/abs/2203.04327} {arXiv:2203.04327
  [hep-ph]} \BibitemShut {NoStop}%
\bibitem [{\citenamefont {Dong}\ \emph {et~al.}(2022)\citenamefont {Dong},
  \citenamefont {Lin},\ and\ \citenamefont {Zou}}]{Dong:2022cuw}%
  \BibitemOpen
  \bibfield  {author} {\bibinfo {author} {\bibfnamefont {Xiang-Kun}\
  \bibnamefont {Dong}}, \bibinfo {author} {\bibfnamefont {Yong-Hui}\
  \bibnamefont {Lin}}, \ and\ \bibinfo {author} {\bibfnamefont {Bing-Song}\
  \bibnamefont {Zou}},\ }\bibfield  {title} {\enquote {\bibinfo {title}
  {{Interpretation of the $\eta_1(1855)$ as a $K\bar{K}_{1}(1400) + c.c.$
  molecule}},}\ }\href {\doibase 10.1007/s11433-022-1887-5} {\bibfield
  {journal} {\bibinfo  {journal} {Sci. China Phys. Mech. Astron.}\ }\textbf
  {\bibinfo {volume} {65}},\ \bibinfo {pages} {261011} (\bibinfo {year}
  {2022})},\ \Eprint {http://arxiv.org/abs/2202.00863} {arXiv:2202.00863
  [hep-ph]} \BibitemShut {NoStop}%
\bibitem [{\citenamefont {Yang}\ \emph {et~al.}(2023)\citenamefont {Yang},
  \citenamefont {Zhu},\ and\ \citenamefont {Huang}}]{Yang:2022rck}%
  \BibitemOpen
  \bibfield  {author} {\bibinfo {author} {\bibfnamefont {Feng}\ \bibnamefont
  {Yang}}, \bibinfo {author} {\bibfnamefont {Hong~Qiang}\ \bibnamefont {Zhu}},
  \ and\ \bibinfo {author} {\bibfnamefont {Yin}\ \bibnamefont {Huang}},\
  }\bibfield  {title} {\enquote {\bibinfo {title} {{Analysis of the
  $\eta_1(1855)$ as a $KK_1(1400)$ molecular state}},}\ }\href {\doibase
  10.1016/j.nuclphysa.2022.122571} {\bibfield  {journal} {\bibinfo  {journal}
  {Nucl. Phys. A}\ }\textbf {\bibinfo {volume} {1030}},\ \bibinfo {pages}
  {122571} (\bibinfo {year} {2023})},\ \Eprint
  {http://arxiv.org/abs/2203.06934} {arXiv:2203.06934 [hep-ph]} \BibitemShut
  {NoStop}%
\bibitem [{\citenamefont {Chen}\ \emph {et~al.}(2008)\citenamefont {Chen},
  \citenamefont {Hosaka},\ and\ \citenamefont {Zhu}}]{Chen:2008ne}%
  \BibitemOpen
  \bibfield  {author} {\bibinfo {author} {\bibfnamefont {Hua-Xing}\
  \bibnamefont {Chen}}, \bibinfo {author} {\bibfnamefont {Atsushi}\
  \bibnamefont {Hosaka}}, \ and\ \bibinfo {author} {\bibfnamefont {Shi-Lin}\
  \bibnamefont {Zhu}},\ }\bibfield  {title} {\enquote {\bibinfo {title} {{$I^G
  J^{PC} = 0^+ 1^{-+}$ tetraquark state}},}\ }\href {\doibase
  10.1103/PhysRevD.78.117502} {\bibfield  {journal} {\bibinfo  {journal} {Phys.
  Rev. D}\ }\textbf {\bibinfo {volume} {78}},\ \bibinfo {pages} {117502}
  (\bibinfo {year} {2008})},\ \Eprint {http://arxiv.org/abs/0808.2344}
  {arXiv:0808.2344 [hep-ph]} \BibitemShut {NoStop}%
\bibitem [{\citenamefont {Wan}\ \emph {et~al.}(2022)\citenamefont {Wan},
  \citenamefont {Zhang},\ and\ \citenamefont {Qiao}}]{Wan:2022xkx}%
  \BibitemOpen
  \bibfield  {author} {\bibinfo {author} {\bibfnamefont {Bing-Dong}\
  \bibnamefont {Wan}}, \bibinfo {author} {\bibfnamefont {Sheng-Qi}\
  \bibnamefont {Zhang}}, \ and\ \bibinfo {author} {\bibfnamefont {Cong-Feng}\
  \bibnamefont {Qiao}},\ }\bibfield  {title} {\enquote {\bibinfo {title} {{A
  possible structure of newly found exotic state $\eta_1(1855)$}},}\
  }\href@noop {} {\  (\bibinfo {year} {2022})},\ \Eprint
  {http://arxiv.org/abs/2203.14014} {arXiv:2203.14014 [hep-ph]} \BibitemShut
  {NoStop}%
\bibitem [{\citenamefont {Jiang}\ \emph {et~al.}(2023)\citenamefont {Jiang},
  \citenamefont {Chen}, \citenamefont {Chen}, \citenamefont {Gong},
  \citenamefont {Li}, \citenamefont {Liu}, \citenamefont {Sun},\ and\
  \citenamefont {Zhang}}]{Jiang:2022gnd}%
  \BibitemOpen
  \bibfield  {author} {\bibinfo {author} {\bibfnamefont {Xiangyu}\ \bibnamefont
  {Jiang}}, \bibinfo {author} {\bibfnamefont {Feiyu}\ \bibnamefont {Chen}},
  \bibinfo {author} {\bibfnamefont {Ying}\ \bibnamefont {Chen}}, \bibinfo
  {author} {\bibfnamefont {Ming}\ \bibnamefont {Gong}}, \bibinfo {author}
  {\bibfnamefont {Ning}\ \bibnamefont {Li}}, \bibinfo {author} {\bibfnamefont
  {Zhaofeng}\ \bibnamefont {Liu}}, \bibinfo {author} {\bibfnamefont {Wei}\
  \bibnamefont {Sun}}, \ and\ \bibinfo {author} {\bibfnamefont {Renqiang}\
  \bibnamefont {Zhang}},\ }\bibfield  {title} {\enquote {\bibinfo {title}
  {{Radiative Decay Width of $J/\psi\to\gamma{\eta}_{(2)}$ from $N_f=2$ Lattice
  QCD}},}\ }\href {\doibase 10.1103/PhysRevLett.130.061901} {\bibfield
  {journal} {\bibinfo  {journal} {Phys. Rev. Lett.}\ }\textbf {\bibinfo
  {volume} {130}},\ \bibinfo {pages} {061901} (\bibinfo {year} {2023})},\
  \Eprint {http://arxiv.org/abs/2206.02724} {arXiv:2206.02724 [hep-lat]}
  \BibitemShut {NoStop}%
\bibitem [{\citenamefont {Gui}\ \emph {et~al.}(2013)\citenamefont {Gui},
  \citenamefont {Chen}, \citenamefont {Li}, \citenamefont {Liu}, \citenamefont
  {Liu}, \citenamefont {Ma}, \citenamefont {Yang},\ and\ \citenamefont
  {Zhang}}]{Gui:2012gx}%
  \BibitemOpen
  \bibfield  {author} {\bibinfo {author} {\bibfnamefont {Long-Cheng}\
  \bibnamefont {Gui}}, \bibinfo {author} {\bibfnamefont {Ying}\ \bibnamefont
  {Chen}}, \bibinfo {author} {\bibfnamefont {Gang}\ \bibnamefont {Li}},
  \bibinfo {author} {\bibfnamefont {Chuan}\ \bibnamefont {Liu}}, \bibinfo
  {author} {\bibfnamefont {Yu-Bin}\ \bibnamefont {Liu}}, \bibinfo {author}
  {\bibfnamefont {Jian-Ping}\ \bibnamefont {Ma}}, \bibinfo {author}
  {\bibfnamefont {Yi-Bo}\ \bibnamefont {Yang}}, \ and\ \bibinfo {author}
  {\bibfnamefont {Jian-Bo}\ \bibnamefont {Zhang}} (\bibinfo {collaboration}
  {CLQCD}),\ }\bibfield  {title} {\enquote {\bibinfo {title} {{Scalar Glueball
  in Radiative $J/\psi$ Decay on the Lattice}},}\ }\href {\doibase
  10.1103/PhysRevLett.110.021601} {\bibfield  {journal} {\bibinfo  {journal}
  {Phys. Rev. Lett.}\ }\textbf {\bibinfo {volume} {110}},\ \bibinfo {pages}
  {021601} (\bibinfo {year} {2013})},\ \Eprint {http://arxiv.org/abs/1206.0125}
  {arXiv:1206.0125 [hep-lat]} \BibitemShut {NoStop}%
\bibitem [{\citenamefont {Yang}\ \emph
  {et~al.}(2013{\natexlab{a}})\citenamefont {Yang}, \citenamefont {Gui},
  \citenamefont {Chen}, \citenamefont {Liu}, \citenamefont {Liu}, \citenamefont
  {Ma},\ and\ \citenamefont {Zhang}}]{Yang:2013xba}%
  \BibitemOpen
  \bibfield  {author} {\bibinfo {author} {\bibfnamefont {Yi-Bo}\ \bibnamefont
  {Yang}}, \bibinfo {author} {\bibfnamefont {Long-Cheng}\ \bibnamefont {Gui}},
  \bibinfo {author} {\bibfnamefont {Ying}\ \bibnamefont {Chen}}, \bibinfo
  {author} {\bibfnamefont {Chuan}\ \bibnamefont {Liu}}, \bibinfo {author}
  {\bibfnamefont {Yu-Bin}\ \bibnamefont {Liu}}, \bibinfo {author}
  {\bibfnamefont {Jian-Ping}\ \bibnamefont {Ma}}, \ and\ \bibinfo {author}
  {\bibfnamefont {Jian-Bo}\ \bibnamefont {Zhang}} (\bibinfo {collaboration}
  {CLQCD}),\ }\bibfield  {title} {\enquote {\bibinfo {title} {{Lattice Study of
  Radiative $J/{\psi}$ Decay to a Tensor Glueball}},}\ }\href {\doibase
  10.1103/PhysRevLett.111.091601} {\bibfield  {journal} {\bibinfo  {journal}
  {Phys. Rev. Lett.}\ }\textbf {\bibinfo {volume} {111}},\ \bibinfo {pages}
  {091601} (\bibinfo {year} {2013}{\natexlab{a}})},\ \Eprint
  {http://arxiv.org/abs/1304.3807} {arXiv:1304.3807 [hep-lat]} \BibitemShut
  {NoStop}%
\bibitem [{\citenamefont {Gui}\ \emph {et~al.}(2019)\citenamefont {Gui},
  \citenamefont {Dong}, \citenamefont {Chen},\ and\ \citenamefont
  {Yang}}]{Gui:2019dtm}%
  \BibitemOpen
  \bibfield  {author} {\bibinfo {author} {\bibfnamefont {Long-Cheng}\
  \bibnamefont {Gui}}, \bibinfo {author} {\bibfnamefont {Jia-Mei}\ \bibnamefont
  {Dong}}, \bibinfo {author} {\bibfnamefont {Ying}\ \bibnamefont {Chen}}, \
  and\ \bibinfo {author} {\bibfnamefont {Yi-Bo}\ \bibnamefont {Yang}},\
  }\bibfield  {title} {\enquote {\bibinfo {title} {{Study of the pseudoscalar
  glueball in $J/\psi$ radiative decays}},}\ }\href {\doibase
  10.1103/PhysRevD.100.054511} {\bibfield  {journal} {\bibinfo  {journal}
  {Phys. Rev. D}\ }\textbf {\bibinfo {volume} {100}},\ \bibinfo {pages}
  {054511} (\bibinfo {year} {2019})},\ \Eprint
  {http://arxiv.org/abs/1906.03666} {arXiv:1906.03666 [hep-lat]} \BibitemShut
  {NoStop}%
\bibitem [{\citenamefont {Peardon}\ \emph {et~al.}(2009)\citenamefont
  {Peardon}, \citenamefont {Bulava}, \citenamefont {Foley}, \citenamefont
  {Morningstar}, \citenamefont {Dudek}, \citenamefont {Edwards}, \citenamefont
  {Joo}, \citenamefont {Lin}, \citenamefont {Richards},\ and\ \citenamefont
  {Juge}}]{Peardon:2009gh}%
  \BibitemOpen
  \bibfield  {author} {\bibinfo {author} {\bibfnamefont {Michael}\ \bibnamefont
  {Peardon}}, \bibinfo {author} {\bibfnamefont {John}\ \bibnamefont {Bulava}},
  \bibinfo {author} {\bibfnamefont {Justin}\ \bibnamefont {Foley}}, \bibinfo
  {author} {\bibfnamefont {Colin}\ \bibnamefont {Morningstar}}, \bibinfo
  {author} {\bibfnamefont {Jozef}\ \bibnamefont {Dudek}}, \bibinfo {author}
  {\bibfnamefont {Robert~G.}\ \bibnamefont {Edwards}}, \bibinfo {author}
  {\bibfnamefont {Balint}\ \bibnamefont {Joo}}, \bibinfo {author}
  {\bibfnamefont {Huey-Wen}\ \bibnamefont {Lin}}, \bibinfo {author}
  {\bibfnamefont {David~G.}\ \bibnamefont {Richards}}, \ and\ \bibinfo {author}
  {\bibfnamefont {Keisuke~Jimmy}\ \bibnamefont {Juge}} (\bibinfo
  {collaboration} {Hadron Spectrum}),\ }\bibfield  {title} {\enquote {\bibinfo
  {title} {{A Novel quark-field creation operator construction for hadronic
  physics in lattice QCD}},}\ }\href {\doibase 10.1103/PhysRevD.80.054506}
  {\bibfield  {journal} {\bibinfo  {journal} {Phys. Rev. D}\ }\textbf {\bibinfo
  {volume} {80}},\ \bibinfo {pages} {054506} (\bibinfo {year} {2009})},\
  \Eprint {http://arxiv.org/abs/0905.2160} {arXiv:0905.2160 [hep-lat]}
  \BibitemShut {NoStop}%
\bibitem [{\citenamefont {Jiang}\ \emph {et~al.}(2022)\citenamefont {Jiang},
  \citenamefont {Sun}, \citenamefont {Chen}, \citenamefont {Chen},
  \citenamefont {Gong}, \citenamefont {Liu},\ and\ \citenamefont
  {Zhang}}]{Jiang:2022ffl}%
  \BibitemOpen
  \bibfield  {author} {\bibinfo {author} {\bibfnamefont {Xiangyu}\ \bibnamefont
  {Jiang}}, \bibinfo {author} {\bibfnamefont {Wei}\ \bibnamefont {Sun}},
  \bibinfo {author} {\bibfnamefont {Feiyu}\ \bibnamefont {Chen}}, \bibinfo
  {author} {\bibfnamefont {Ying}\ \bibnamefont {Chen}}, \bibinfo {author}
  {\bibfnamefont {Ming}\ \bibnamefont {Gong}}, \bibinfo {author} {\bibfnamefont
  {Zhaofeng}\ \bibnamefont {Liu}}, \ and\ \bibinfo {author} {\bibfnamefont
  {Renqiang}\ \bibnamefont {Zhang}},\ }\bibfield  {title} {\enquote {\bibinfo
  {title} {{$\eta$-glueball mixing from $N_f=2$ lattice QCD}},}\ }\href@noop {}
  {\  (\bibinfo {year} {2022})},\ \Eprint {http://arxiv.org/abs/2205.12541}
  {arXiv:2205.12541 [hep-lat]} \BibitemShut {NoStop}%
\bibitem [{\citenamefont {Meng}\ \emph {et~al.}(2009)\citenamefont {Meng} \emph
  {et~al.}}]{CLQCD:2009nvn}%
  \BibitemOpen
  \bibfield  {author} {\bibinfo {author} {\bibfnamefont {Guo-Zhan}\
  \bibnamefont {Meng}} \emph {et~al.} (\bibinfo {collaboration} {CLQCD}),\
  }\bibfield  {title} {\enquote {\bibinfo {title} {{Low-energy $D^{*+}
  \bar{D}^0_1$ Scattering and the Resonance-like Structure $Z^+(4430)$}},}\
  }\href {\doibase 10.1103/PhysRevD.80.034503} {\bibfield  {journal} {\bibinfo
  {journal} {Phys. Rev. D}\ }\textbf {\bibinfo {volume} {80}},\ \bibinfo
  {pages} {034503} (\bibinfo {year} {2009})},\ \Eprint
  {http://arxiv.org/abs/0905.0752} {arXiv:0905.0752 [hep-lat]} \BibitemShut
  {NoStop}%
\bibitem [{\citenamefont {Dudek}\ \emph {et~al.}(2006)\citenamefont {Dudek},
  \citenamefont {Edwards},\ and\ \citenamefont {Richards}}]{Dudek:2006ej}%
  \BibitemOpen
  \bibfield  {author} {\bibinfo {author} {\bibfnamefont {Jozef~J.}\
  \bibnamefont {Dudek}}, \bibinfo {author} {\bibfnamefont {Robert~G.}\
  \bibnamefont {Edwards}}, \ and\ \bibinfo {author} {\bibfnamefont {David~G.}\
  \bibnamefont {Richards}},\ }\bibfield  {title} {\enquote {\bibinfo {title}
  {{Radiative transitions in charmonium from lattice QCD}},}\ }\href {\doibase
  10.1103/PhysRevD.73.074507} {\bibfield  {journal} {\bibinfo  {journal} {Phys.
  Rev. D}\ }\textbf {\bibinfo {volume} {73}},\ \bibinfo {pages} {074507}
  (\bibinfo {year} {2006})},\ \Eprint {http://arxiv.org/abs/hep-ph/0601137}
  {arXiv:hep-ph/0601137} \BibitemShut {NoStop}%
\bibitem [{\citenamefont {Dudek}\ \emph {et~al.}(2009)\citenamefont {Dudek},
  \citenamefont {Edwards},\ and\ \citenamefont {Thomas}}]{Dudek:2009kk}%
  \BibitemOpen
  \bibfield  {author} {\bibinfo {author} {\bibfnamefont {Jozef~J.}\
  \bibnamefont {Dudek}}, \bibinfo {author} {\bibfnamefont {Robert}\
  \bibnamefont {Edwards}}, \ and\ \bibinfo {author} {\bibfnamefont
  {Christopher~E.}\ \bibnamefont {Thomas}},\ }\bibfield  {title} {\enquote
  {\bibinfo {title} {{Exotic and excited-state radiative transitions in
  charmonium from lattice QCD}},}\ }\href {\doibase 10.1103/PhysRevD.79.094504}
  {\bibfield  {journal} {\bibinfo  {journal} {Phys. Rev. D}\ }\textbf {\bibinfo
  {volume} {79}},\ \bibinfo {pages} {094504} (\bibinfo {year} {2009})},\
  \Eprint {http://arxiv.org/abs/0902.2241} {arXiv:0902.2241 [hep-ph]}
  \BibitemShut {NoStop}%
\bibitem [{\citenamefont {Thomas}\ \emph {et~al.}(2012)\citenamefont {Thomas},
  \citenamefont {Edwards},\ and\ \citenamefont {Dudek}}]{Thomas:2011rh}%
  \BibitemOpen
  \bibfield  {author} {\bibinfo {author} {\bibfnamefont {Christopher~E.}\
  \bibnamefont {Thomas}}, \bibinfo {author} {\bibfnamefont {Robert~G.}\
  \bibnamefont {Edwards}}, \ and\ \bibinfo {author} {\bibfnamefont {Jozef~J.}\
  \bibnamefont {Dudek}},\ }\bibfield  {title} {\enquote {\bibinfo {title}
  {{Helicity operators for mesons in flight on the lattice}},}\ }\href
  {\doibase 10.1103/PhysRevD.85.014507} {\bibfield  {journal} {\bibinfo
  {journal} {Phys. Rev. D}\ }\textbf {\bibinfo {volume} {85}},\ \bibinfo
  {pages} {014507} (\bibinfo {year} {2012})},\ \Eprint
  {http://arxiv.org/abs/1107.1930} {arXiv:1107.1930 [hep-lat]} \BibitemShut
  {NoStop}%
\bibitem [{\citenamefont {Bali}\ \emph {et~al.}(2016)\citenamefont {Bali},
  \citenamefont {Lang}, \citenamefont {Musch},\ and\ \citenamefont
  {Sch\"afer}}]{Bali:2016lva}%
  \BibitemOpen
  \bibfield  {author} {\bibinfo {author} {\bibfnamefont {Gunnar~S.}\
  \bibnamefont {Bali}}, \bibinfo {author} {\bibfnamefont {Bernhard}\
  \bibnamefont {Lang}}, \bibinfo {author} {\bibfnamefont {Bernhard~U.}\
  \bibnamefont {Musch}}, \ and\ \bibinfo {author} {\bibfnamefont {Andreas}\
  \bibnamefont {Sch\"afer}},\ }\bibfield  {title} {\enquote {\bibinfo {title}
  {{Novel quark smearing for hadrons with high momenta in lattice QCD}},}\
  }\href {\doibase 10.1103/PhysRevD.93.094515} {\bibfield  {journal} {\bibinfo
  {journal} {Phys. Rev. D}\ }\textbf {\bibinfo {volume} {93}},\ \bibinfo
  {pages} {094515} (\bibinfo {year} {2016})},\ \Eprint
  {http://arxiv.org/abs/1602.05525} {arXiv:1602.05525 [hep-lat]} \BibitemShut
  {NoStop}%
\bibitem [{\citenamefont {Ma}\ \emph {et~al.}(2021{\natexlab{a}})\citenamefont
  {Ma}, \citenamefont {Chen}, \citenamefont {Gong},\ and\ \citenamefont
  {Liu}}]{Ma:2020bex}%
  \BibitemOpen
  \bibfield  {author} {\bibinfo {author} {\bibfnamefont {Yunheng}\ \bibnamefont
  {Ma}}, \bibinfo {author} {\bibfnamefont {Ying}\ \bibnamefont {Chen}},
  \bibinfo {author} {\bibfnamefont {Ming}\ \bibnamefont {Gong}}, \ and\
  \bibinfo {author} {\bibfnamefont {Zhaofeng}\ \bibnamefont {Liu}},\ }\bibfield
   {title} {\enquote {\bibinfo {title} {{Strangeonium-like hybrids on the
  lattice}},}\ }\href {\doibase 10.1088/1674-1137/abc241} {\bibfield  {journal}
  {\bibinfo  {journal} {Chin. Phys. C}\ }\textbf {\bibinfo {volume} {45}},\
  \bibinfo {pages} {013112} (\bibinfo {year} {2021}{\natexlab{a}})},\ \Eprint
  {http://arxiv.org/abs/2007.14893} {arXiv:2007.14893 [hep-lat]} \BibitemShut
  {NoStop}%
\bibitem [{\citenamefont {Ma}\ \emph {et~al.}(2021{\natexlab{b}})\citenamefont
  {Ma}, \citenamefont {Sun}, \citenamefont {Chen}, \citenamefont {Gong},\ and\
  \citenamefont {Liu}}]{Ma:2019hsm}%
  \BibitemOpen
  \bibfield  {author} {\bibinfo {author} {\bibfnamefont {Yunheng}\ \bibnamefont
  {Ma}}, \bibinfo {author} {\bibfnamefont {Wei}\ \bibnamefont {Sun}}, \bibinfo
  {author} {\bibfnamefont {Ying}\ \bibnamefont {Chen}}, \bibinfo {author}
  {\bibfnamefont {Ming}\ \bibnamefont {Gong}}, \ and\ \bibinfo {author}
  {\bibfnamefont {Zhaofeng}\ \bibnamefont {Liu}},\ }\bibfield  {title}
  {\enquote {\bibinfo {title} {{Color halo scenario of charmonium-like
  hybrids}},}\ }\href {\doibase 10.1088/1674-1137/ac0ee2} {\bibfield  {journal}
  {\bibinfo  {journal} {Chin. Phys. C}\ }\textbf {\bibinfo {volume} {45}},\
  \bibinfo {pages} {093111} (\bibinfo {year} {2021}{\natexlab{b}})},\ \Eprint
  {http://arxiv.org/abs/1910.09819} {arXiv:1910.09819 [hep-lat]} \BibitemShut
  {NoStop}%
\bibitem [{\citenamefont {Yang}\ \emph
  {et~al.}(2013{\natexlab{b}})\citenamefont {Yang}, \citenamefont {Chen},
  \citenamefont {Gui}, \citenamefont {Liu}, \citenamefont {Liu}, \citenamefont
  {Liu}, \citenamefont {Ma},\ and\ \citenamefont {Zhang}}]{Yang:2012mya}%
  \BibitemOpen
  \bibfield  {author} {\bibinfo {author} {\bibfnamefont {Yi-Bo}\ \bibnamefont
  {Yang}}, \bibinfo {author} {\bibfnamefont {Ying}\ \bibnamefont {Chen}},
  \bibinfo {author} {\bibfnamefont {Long-Cheng}\ \bibnamefont {Gui}}, \bibinfo
  {author} {\bibfnamefont {Chuan}\ \bibnamefont {Liu}}, \bibinfo {author}
  {\bibfnamefont {Yu-Bin}\ \bibnamefont {Liu}}, \bibinfo {author}
  {\bibfnamefont {Zhaofeng}\ \bibnamefont {Liu}}, \bibinfo {author}
  {\bibfnamefont {Jian-Ping}\ \bibnamefont {Ma}}, \ and\ \bibinfo {author}
  {\bibfnamefont {Jian-Bo}\ \bibnamefont {Zhang}} (\bibinfo {collaboration}
  {CLQCD}),\ }\bibfield  {title} {\enquote {\bibinfo {title} {{Lattice study on
  $\eta_{c2}$ and X(3872)}},}\ }\href {\doibase 10.1103/PhysRevD.87.014501}
  {\bibfield  {journal} {\bibinfo  {journal} {Phys. Rev. D}\ }\textbf {\bibinfo
  {volume} {87}},\ \bibinfo {pages} {014501} (\bibinfo {year}
  {2013}{\natexlab{b}})},\ \Eprint {http://arxiv.org/abs/1206.2086}
  {arXiv:1206.2086 [hep-lat]} \BibitemShut {NoStop}%
\bibitem [{\citenamefont {Brice\~no}\ and\ \citenamefont
  {Hansen}(2015)}]{Briceno:2015csa}%
  \BibitemOpen
  \bibfield  {author} {\bibinfo {author} {\bibfnamefont {Ra\'ul~A.}\
  \bibnamefont {Brice\~no}}\ and\ \bibinfo {author} {\bibfnamefont
  {Maxwell~T.}\ \bibnamefont {Hansen}},\ }\bibfield  {title} {\enquote
  {\bibinfo {title} {{Multichannel 0 $\to$ 2 and 1 $\to$ 2 transition
  amplitudes for arbitrary spin particles in a finite volume}},}\ }\href
  {\doibase 10.1103/PhysRevD.92.074509} {\bibfield  {journal} {\bibinfo
  {journal} {Phys. Rev. D}\ }\textbf {\bibinfo {volume} {92}},\ \bibinfo
  {pages} {074509} (\bibinfo {year} {2015})},\ \Eprint
  {http://arxiv.org/abs/1502.04314} {arXiv:1502.04314 [hep-lat]} \BibitemShut
  {NoStop}%
\bibitem [{\citenamefont {Brice\~no}\ \emph {et~al.}(2016)\citenamefont
  {Brice\~no}, \citenamefont {Dudek}, \citenamefont {Edwards}, \citenamefont
  {Shultz}, \citenamefont {Thomas},\ and\ \citenamefont
  {Wilson}}]{Briceno:2016kkp}%
  \BibitemOpen
  \bibfield  {author} {\bibinfo {author} {\bibfnamefont {Ra\'ul~A.}\
  \bibnamefont {Brice\~no}}, \bibinfo {author} {\bibfnamefont {Jozef~J.}\
  \bibnamefont {Dudek}}, \bibinfo {author} {\bibfnamefont {Robert~G.}\
  \bibnamefont {Edwards}}, \bibinfo {author} {\bibfnamefont {Christian~J.}\
  \bibnamefont {Shultz}}, \bibinfo {author} {\bibfnamefont {Christopher~E.}\
  \bibnamefont {Thomas}}, \ and\ \bibinfo {author} {\bibfnamefont {David~J.}\
  \bibnamefont {Wilson}},\ }\bibfield  {title} {\enquote {\bibinfo {title}
  {{The $\pi\pi\to\pi\gamma^\star$ amplitude and the resonant
  $\rho\to\pi\gamma^\star$ transition from lattice QCD}},}\ }\href {\doibase
  10.1103/PhysRevD.93.114508} {\bibfield  {journal} {\bibinfo  {journal} {Phys.
  Rev. D}\ }\textbf {\bibinfo {volume} {93}},\ \bibinfo {pages} {114508}
  (\bibinfo {year} {2016})},\ \bibinfo {note} {[Erratum: Phys.Rev.D 105, 079902
  (2022)]},\ \Eprint {http://arxiv.org/abs/1604.03530} {arXiv:1604.03530
  [hep-ph]} \BibitemShut {NoStop}%
\bibitem [{\citenamefont {Brice\~no}\ \emph {et~al.}(2018)\citenamefont
  {Brice\~no}, \citenamefont {Dudek},\ and\ \citenamefont
  {Young}}]{Briceno:2017max}%
  \BibitemOpen
  \bibfield  {author} {\bibinfo {author} {\bibfnamefont {Ra\'ul~A.}\
  \bibnamefont {Brice\~no}}, \bibinfo {author} {\bibfnamefont {Jozef~J.}\
  \bibnamefont {Dudek}}, \ and\ \bibinfo {author} {\bibfnamefont {Ross~D.}\
  \bibnamefont {Young}},\ }\bibfield  {title} {\enquote {\bibinfo {title}
  {{Scattering processes and resonances from lattice QCD}},}\ }\href {\doibase
  10.1103/RevModPhys.90.025001} {\bibfield  {journal} {\bibinfo  {journal}
  {Rev. Mod. Phys.}\ }\textbf {\bibinfo {volume} {90}},\ \bibinfo {pages}
  {025001} (\bibinfo {year} {2018})},\ \Eprint
  {http://arxiv.org/abs/1706.06223} {arXiv:1706.06223 [hep-lat]} \BibitemShut
  {NoStop}%
\bibitem [{\citenamefont {Alexandrou}\ \emph {et~al.}(2018)\citenamefont
  {Alexandrou}, \citenamefont {Leskovec}, \citenamefont {Meinel}, \citenamefont
  {Negele}, \citenamefont {Paul}, \citenamefont {Petschlies}, \citenamefont
  {Pochinsky}, \citenamefont {Rendon},\ and\ \citenamefont
  {Syritsyn}}]{Alexandrou:2018jbt}%
  \BibitemOpen
  \bibfield  {author} {\bibinfo {author} {\bibfnamefont {Constantia}\
  \bibnamefont {Alexandrou}}, \bibinfo {author} {\bibfnamefont {Luka}\
  \bibnamefont {Leskovec}}, \bibinfo {author} {\bibfnamefont {Stefan}\
  \bibnamefont {Meinel}}, \bibinfo {author} {\bibfnamefont {John}\ \bibnamefont
  {Negele}}, \bibinfo {author} {\bibfnamefont {Srijit}\ \bibnamefont {Paul}},
  \bibinfo {author} {\bibfnamefont {Marcus}\ \bibnamefont {Petschlies}},
  \bibinfo {author} {\bibfnamefont {Andrew}\ \bibnamefont {Pochinsky}},
  \bibinfo {author} {\bibfnamefont {Gumaro}\ \bibnamefont {Rendon}}, \ and\
  \bibinfo {author} {\bibfnamefont {Sergey}\ \bibnamefont {Syritsyn}},\
  }\bibfield  {title} {\enquote {\bibinfo {title} {{$\pi\gamma \to \pi\pi$
  transition and the $\rho$ radiative decay width from lattice QCD}},}\ }\href
  {\doibase 10.1103/PhysRevD.98.074502} {\bibfield  {journal} {\bibinfo
  {journal} {Phys. Rev. D}\ }\textbf {\bibinfo {volume} {98}},\ \bibinfo
  {pages} {074502} (\bibinfo {year} {2018})},\ \bibinfo {note} {[Erratum:
  Phys.Rev.D 105, 019902 (2022)]},\ \Eprint {http://arxiv.org/abs/1807.08357}
  {arXiv:1807.08357 [hep-lat]} \BibitemShut {NoStop}%
\bibitem [{\citenamefont {Brice\~no}\ \emph {et~al.}(2021)\citenamefont
  {Brice\~no}, \citenamefont {Dudek},\ and\ \citenamefont
  {Leskovec}}]{Briceno:2021xlc}%
  \BibitemOpen
  \bibfield  {author} {\bibinfo {author} {\bibfnamefont {Ra\'ul~A.}\
  \bibnamefont {Brice\~no}}, \bibinfo {author} {\bibfnamefont {Jozef~J.}\
  \bibnamefont {Dudek}}, \ and\ \bibinfo {author} {\bibfnamefont {Luka}\
  \bibnamefont {Leskovec}},\ }\bibfield  {title} {\enquote {\bibinfo {title}
  {{Constraining $1+\mathcal{J}\to 2$ coupled-channel amplitudes in
  finite-volume}},}\ }\href {\doibase 10.1103/PhysRevD.104.054509} {\bibfield
  {journal} {\bibinfo  {journal} {Phys. Rev. D}\ }\textbf {\bibinfo {volume}
  {104}},\ \bibinfo {pages} {054509} (\bibinfo {year} {2021})},\ \Eprint
  {http://arxiv.org/abs/2105.02017} {arXiv:2105.02017 [hep-lat]} \BibitemShut
  {NoStop}%
\bibitem [{\citenamefont {Radhakrishnan}\ \emph {et~al.}(2022)\citenamefont
  {Radhakrishnan}, \citenamefont {Dudek},\ and\ \citenamefont
  {Edwards}}]{Radhakrishnan:2022ubg}%
  \BibitemOpen
  \bibfield  {author} {\bibinfo {author} {\bibfnamefont {Archana}\ \bibnamefont
  {Radhakrishnan}}, \bibinfo {author} {\bibfnamefont {Jozef~J.}\ \bibnamefont
  {Dudek}}, \ and\ \bibinfo {author} {\bibfnamefont {Robert~G.}\ \bibnamefont
  {Edwards}} (\bibinfo {collaboration} {Hadron Spectrum}),\ }\bibfield  {title}
  {\enquote {\bibinfo {title} {{Radiative decay of the resonant $K^*$ and the
  \ensuremath{\gamma}K\textrightarrow{}K\ensuremath{\pi} amplitude from lattice
  QCD}},}\ }\href {\doibase 10.1103/PhysRevD.106.114513} {\bibfield  {journal}
  {\bibinfo  {journal} {Phys. Rev. D}\ }\textbf {\bibinfo {volume} {106}},\
  \bibinfo {pages} {114513} (\bibinfo {year} {2022})},\ \Eprint
  {http://arxiv.org/abs/2208.13755} {arXiv:2208.13755 [hep-lat]} \BibitemShut
  {NoStop}%
\bibitem [{\citenamefont {Zyla}\ \emph {et~al.}(2020)\citenamefont {Zyla} \emph
  {et~al.}}]{Zyla:2020zbs}%
  \BibitemOpen
  \bibfield  {author} {\bibinfo {author} {\bibfnamefont {P.A.}\ \bibnamefont
  {Zyla}} \emph {et~al.} (\bibinfo {collaboration} {Particle Data Group}),\
  }\bibfield  {title} {\enquote {\bibinfo {title} {{Review of Particle
  Physics}},}\ }\href {\doibase 10.1093/ptep/ptaa104} {\bibfield  {journal}
  {\bibinfo  {journal} {PTEP}\ }\textbf {\bibinfo {volume} {2020}},\ \bibinfo
  {pages} {083C01} (\bibinfo {year} {2020})}\BibitemShut {NoStop}%
\bibitem [{\citenamefont {Page}(1997)}]{Page:1996rj}%
  \BibitemOpen
  \bibfield  {author} {\bibinfo {author} {\bibfnamefont {Philip~R.}\
  \bibnamefont {Page}},\ }\bibfield  {title} {\enquote {\bibinfo {title} {{Why
  hybrid meson coupling to two S wave mesons is suppressed}},}\ }\href
  {\doibase 10.1016/S0370-2693(97)00438-3} {\bibfield  {journal} {\bibinfo
  {journal} {Phys. Lett. B}\ }\textbf {\bibinfo {volume} {402}},\ \bibinfo
  {pages} {183--188} (\bibinfo {year} {1997})},\ \Eprint
  {http://arxiv.org/abs/hep-ph/9611375} {arXiv:hep-ph/9611375} \BibitemShut
  {NoStop}%
\bibitem [{\citenamefont {Page}\ \emph {et~al.}(1999)\citenamefont {Page},
  \citenamefont {Swanson},\ and\ \citenamefont {Szczepaniak}}]{Page:1998gz}%
  \BibitemOpen
  \bibfield  {author} {\bibinfo {author} {\bibfnamefont {Philip~R.}\
  \bibnamefont {Page}}, \bibinfo {author} {\bibfnamefont {Eric~S.}\
  \bibnamefont {Swanson}}, \ and\ \bibinfo {author} {\bibfnamefont {Adam~P.}\
  \bibnamefont {Szczepaniak}},\ }\bibfield  {title} {\enquote {\bibinfo {title}
  {{Hybrid meson decay phenomenology}},}\ }\href {\doibase
  10.1103/PhysRevD.59.034016} {\bibfield  {journal} {\bibinfo  {journal} {Phys.
  Rev. D}\ }\textbf {\bibinfo {volume} {59}},\ \bibinfo {pages} {034016}
  (\bibinfo {year} {1999})},\ \Eprint {http://arxiv.org/abs/hep-ph/9808346}
  {arXiv:hep-ph/9808346} \BibitemShut {NoStop}%
\bibitem [{\citenamefont {Edwards}\ and\ \citenamefont
  {Joo}(2005)}]{Edwards:2004sx}%
  \BibitemOpen
  \bibfield  {author} {\bibinfo {author} {\bibfnamefont {Robert~G.}\
  \bibnamefont {Edwards}}\ and\ \bibinfo {author} {\bibfnamefont {Balint}\
  \bibnamefont {Joo}} (\bibinfo {collaboration} {SciDAC, LHPC, UKQCD}),\
  }\bibfield  {title} {\enquote {\bibinfo {title} {{The Chroma software system
  for lattice QCD}},}\ }\href {\doibase 10.1016/j.nuclphysbps.2004.11.254}
  {\bibfield  {journal} {\bibinfo  {journal} {Nucl. Phys. B Proc. Suppl.}\
  }\textbf {\bibinfo {volume} {140}},\ \bibinfo {pages} {832} (\bibinfo {year}
  {2005})},\ \Eprint {http://arxiv.org/abs/hep-lat/0409003}
  {arXiv:hep-lat/0409003} \BibitemShut {NoStop}%
\bibitem [{\citenamefont {Clark}\ \emph {et~al.}(2010)\citenamefont {Clark},
  \citenamefont {Babich}, \citenamefont {Barros}, \citenamefont {Brower},\ and\
  \citenamefont {Rebbi}}]{Clark:2009wm}%
  \BibitemOpen
  \bibfield  {author} {\bibinfo {author} {\bibfnamefont {M.~A.}\ \bibnamefont
  {Clark}}, \bibinfo {author} {\bibfnamefont {R.}~\bibnamefont {Babich}},
  \bibinfo {author} {\bibfnamefont {K.}~\bibnamefont {Barros}}, \bibinfo
  {author} {\bibfnamefont {R.~C.}\ \bibnamefont {Brower}}, \ and\ \bibinfo
  {author} {\bibfnamefont {C.}~\bibnamefont {Rebbi}},\ }\bibfield  {title}
  {\enquote {\bibinfo {title} {{Solving Lattice QCD systems of equations using
  mixed precision solvers on GPUs}},}\ }\href {\doibase
  10.1016/j.cpc.2010.05.002} {\bibfield  {journal} {\bibinfo  {journal}
  {Comput. Phys. Commun.}\ }\textbf {\bibinfo {volume} {181}},\ \bibinfo
  {pages} {1517--1528} (\bibinfo {year} {2010})},\ \Eprint
  {http://arxiv.org/abs/0911.3191} {arXiv:0911.3191 [hep-lat]} \BibitemShut
  {NoStop}%
\bibitem [{\citenamefont {Babich}\ \emph {et~al.}(2011)\citenamefont {Babich},
  \citenamefont {Clark}, \citenamefont {Joo}, \citenamefont {Shi},
  \citenamefont {Brower},\ and\ \citenamefont {Gottlieb}}]{Babich:2011np}%
  \BibitemOpen
  \bibfield  {author} {\bibinfo {author} {\bibfnamefont {R.}~\bibnamefont
  {Babich}}, \bibinfo {author} {\bibfnamefont {M.~A.}\ \bibnamefont {Clark}},
  \bibinfo {author} {\bibfnamefont {B.}~\bibnamefont {Joo}}, \bibinfo {author}
  {\bibfnamefont {G.}~\bibnamefont {Shi}}, \bibinfo {author} {\bibfnamefont
  {R.~C.}\ \bibnamefont {Brower}}, \ and\ \bibinfo {author} {\bibfnamefont
  {S.}~\bibnamefont {Gottlieb}},\ }\bibfield  {title} {\enquote {\bibinfo
  {title} {{Scaling Lattice QCD beyond 100 GPUs}},}\ }in\ \href {\doibase
  10.1145/2063384.2063478} {\emph {\bibinfo {booktitle} {{SC11 International
  Conference for High Performance Computing, Networking, Storage and
  Analysis}}}}\ (\bibinfo {year} {2011})\ \Eprint
  {http://arxiv.org/abs/1109.2935} {arXiv:1109.2935 [hep-lat]} \BibitemShut
  {NoStop}%
\end{thebibliography}%

\appendix
\section{Form factors}
\label{sec:form-factors}
Since the quantum numbers $J^P$ of $\eta_1$ and $J/\psi$ are all $1^-$, the transition matrix $\braket{\eta_1|j^\mu_{\mathrm{em}}|J/\psi}$ is given by the vector-to-vector one $\braket{V|j^\mu_\mathrm{em}|V}$, which can be expanded in terms of form factors by enumerating all possible Lorentz structures

\begin{eqnarray}\label{eq:sp:form_factor_expand}
    &&\braket{V(p', \epsilon')|j^\mu_\mathrm{em}|V(p, \epsilon)} = \nonumber\\
    &&G_1(Q^2) p\cdot\epsilon'^\ast\epsilon^\mu + G_2(Q^2) p'\cdot\epsilon\epsilon'^{\ast\mu}\nonumber\\
    &&+\epsilon\cdot\epsilon'^\ast [G_3(Q^2)(p^\mu + p'^\mu) + G_4(Q^2) q^\mu]\nonumber\\
    &&+ (p\cdot\epsilon')(p'\cdot\epsilon)[G_5(Q^2)(p^\mu + p'^\mu) + G_6(Q^2) q^\mu]
\end{eqnarray}
$G_4(Q^2)$, $G_6(Q^2)$ can be eliminated and expressed in terms of other form factors using the conservation of current $\braket{V|j^\mu_{\mathrm{em}}|V} q_\mu = 0$ as
\begin{eqnarray}
    G_4(Q^2) & = & -\frac{m^2 - m'^2}{q^2}G_3(Q^2)\nonumber\\
    G_6(Q^2) & = & -\frac1{q^2}[G_2(Q^2) - G_1(Q^2) + G_5(Q^2)(m_i^2 - m_f^2)]\nonumber\\
\end{eqnarray}
As in Ref.~[29] of the main article, it is convenient to expand the helicity amplitudes in terms of \textit{multipoles}. In the frame where the initial state is at rest and the photon goes in the $z-$ direction, the amplitudes are
\begin{eqnarray}
    \braket{V^\mp|j^\mu_{\mathrm{em}}|V^0} \epsilon_{\gamma, \mu}^{\pm,\ast} & = &\frac1{\sqrt2}[M_1(Q^2) + E_2(Q^2)]\nonumber\\
    \braket{V^0|j^\mu_{\mathrm{em}}|V^\pm} \epsilon_{\gamma, \mu}^{\pm,\ast} & = &\frac1{\sqrt2}[M_1(Q^2) - E_2(Q^2)]\nonumber\\
    \braket{V^0|j^\mu_{\mathrm{em}}|V^0} \epsilon_{\gamma, \mu}^{0,\ast} & = &\frac1{\sqrt3}C_0(Q^2) - \sqrt{\frac23} C_2(Q^2)\nonumber\\
    \braket{V^\pm|j^\mu_{\mathrm{em}}|V^\pm}\epsilon_{\gamma,\mu}^{0,\ast} & = &\frac1{\sqrt3}C_0(Q^2) + \frac1{\sqrt6}C_2(Q^2),
\end{eqnarray}
where the superscripts $\mp,\pm, 0$ refer the different polarizations of the two vector mesons. On the other hand, these amplitudes can also be expressed in terms of form factors $G_i(Q^2)$ by substituting specific momenta and  polarization vectors into Eq.~\eqref{eq:sp:form_factor_expand}, giving us four equations. By solving these equations the form factors $G_i(Q^2)$ can be related to multipoles $M_1(Q^2)$, $E_2(Q^2)$, $C_0(Q^2)$, $C_2(Q^2)$ as
\begin{eqnarray}\label{eq:sp:two-types}
    G_1(Q^2) &=& \frac{m'}{\sqrt{2\Omega}}(M_1(Q^2) - E_2(Q^2))\nonumber\\
    G_2(Q^2) &=& \frac{m}{\sqrt{2\Omega}}(M_1(Q^2) + E_2(Q^2)) \nonumber\\
    G_3(Q^2) &=& -\frac{\sqrt{q^2}}{4\sqrt{3\Omega}}(2C_0(Q^2) + \sqrt2 C_2(Q^2))\nonumber\\
    G_5(Q^2) &=& \frac1{12\sqrt2\Omega^{3/2}}\left[\sqrt{6q^2} \left((m - m')^2q^2\right)C_0(Q^2) \right.\nonumber\\
        &+& \sqrt{3q^2} \left((m + m')^2 + 2m m' - q^2\right)C_2(Q^2)\nonumber\\
        &+& 3(m' - m)\left((m' + m)^2 - q^2\right)E_2(Q^2)\nonumber\\
        &-& \left. 3(m' + m)\left((m' - m)^2 - q^2\right)M_1(Q^2) \right],
\end{eqnarray}
where $Q^2 = -q^2$, and
\begin{eqnarray}
    \Omega &=& (p\cdot p')^2 - m^2 m'^2 \nonumber\\
           &=& \frac14 [(m + m')^2 + Q^2][(m - m')^2 + Q^2],\label{eq:omega-fn}
\end{eqnarray}
Note that in our case $m = m_{J/\psi}$ and $m' = m_{\eta_1}$. 

\end{document}